\documentclass[a4paper,twocolumn,10pt,accepted=2025-09-02]{quantumarticle}
\pdfoutput=1 
\usepackage{IEEEtrantools}
\usepackage{graphicx}
\usepackage{amssymb}
\usepackage{amsmath}
\usepackage{braket}
\usepackage{physics}
\usepackage{hyperref}
\usepackage{footnote}
\usepackage{xcolor}
\usepackage{multirow}
\usepackage{enumerate}
\usepackage{capt-of}
\usepackage{subfigure}
\usepackage{soul}

\newcommand{\SO}[1]{\mathcal{#1}}
\newcommand{\mat}[1]{\mathbb{#1}}
\newcommand{\kket}[1]{\kern-0.2em\left.\ket{#1}\kern-0.18em\right\rangle}
\newcommand{\bbra}[1]{\left\langle\kern-0.18em\bra{#1}\right.\kern-0.2em}
\newcommand{\SOev}[2]{\left\langle\kern-0.18em\braket{#1}{#2}\kern-0.18em\right\rangle}

\newcommand{\sub}{\IEEEyessubnumber}
\newcommand{\startsub}{\IEEEyesnumber\IEEEyessubnumber}
\newcommand{\T}{\intercal}
\renewcommand{\~}[1]{\Tilde{#1}}
\renewcommand{\vdot}{{}_{{}^\bullet}}

\date{2 Spetember 2025}

\begin{document}
\title{Quantifying spectral signatures of non-Markovianity beyond the Born-Redfield master equation}
\author{A. Keefe}
\email{andrew\_keefe@uml.edu}
\affiliation{Department of Physics and Applied Physics, University of Massachusetts, Lowell, MA 01854, USA}
\author{N. Agarwal}
\email{nishant\_agarwal@uml.edu}
\affiliation{Department of Physics and Applied Physics, University of Massachusetts, Lowell, MA 01854, USA}
\author{A. Kamal}
\email{archana.kamal@northwestern.edu}
\affiliation{Department of Physics and Applied Physics, University of Massachusetts, Lowell, MA 01854, USA}
\affiliation{Department of Physics and Astronomy, Northwestern University, Evanston, IL 60208, USA}

\begin{abstract}
Memory or time-non-local effects in open quantum dynamics pose theoretical as well as practical challenges in the understanding and control of noisy quantum systems. While there has been a comprehensive and concerted effort towards developing diagnostics for non-Markovian dynamics, all existing measures rely on time-domain measurements which are typically slow and expensive as they require averaging several runs to resolve small transient features on a broad background, and scale unfavorably with system size and complexity. In this work, we propose a spectroscopic measure of non-Markovianity which can detect persistent non-Markovianity in the system steady state. In addition to being experimentally viable, the proposed measure has a direct information theoretic interpretation: a large value indicates the information loss per unit bandwidth of making the Markov approximation. In the same vein, we derive a frequency-domain quantum master equation (FD-QME) that goes beyond the standard Born-Redfield description and retains the full memory of the state of the reduced system. Using the FD-QME and the proposed measure, we are able to reliably diagnose and quantify non-Markovianity in several system-bath settings including those with bath correlations and retardation effects.
\end{abstract}
\maketitle
%


%
\section{Introduction}
\label{sec:Introduction}
%
%
Quantum master equations (QMEs) are universal tools to describe the dynamics of open quantum systems, be it to capture undesired decoherence due to loss of information and energy to uncontrolled degrees of freedom or, more recently, to investigate bespoke dissipation as a resource for implementing quantum error correction. In developing effective QME descriptions, several approximations are commonly employed in order to obtain a tractable equation of motion for the density operator describing the reduced system(s) of interest. One of the most popular QME is the GKSL (for Gorini–Kossakowski–Sudarshan–Lindblad) master equation, frequently abbreviated as the ``Lindblad QME'', which generates a quantum dynamical semigroup for the dynamics of the density operator. Originally proposed in the context of quantum optical setups, its simplicity of form and tractability, in terms of enabling closed-form, analytical solutions, have made Lindblad QME ubiquitous in the theory of open systems.
\par
The general scheme of approximations followed to obtain such a QME has a Born-Oppenheimer flavor to it: first weak coupling, or an `entanglement breaking' ansatz, is imposed between system-bath states at all times enabling a perturbative expansion, followed by a separation of time scales, or `quasi-static bath' ansatz, enabling an adiabatic elimination of fast bath degrees of freedom \cite{Manzano2020}. Interestingly, while these may seem a bit extreme at first, such approximations turn out to be well-motivated in a surprisingly large number of scenarios where the bath consists of many uncorrelated degrees of freedom. A canonical example is the famous Caldeira-Leggett model \cite{Caldeira1981,Caldeira1982}, defined as a thermal bath consisting of several harmonic oscillators, that is routinely employed in microscopic derivation of quantum Brownian motion. Similarly, under the rotating wave approximation, where the system samples only a narrow density of states of the bath near its resonance, an effective `white' bath spectral density -- reminiscent of that obtained for fast-evolving uncorrelated modes -- can be assumed, leading to a Markovian QME for the system. 
\par
%
Any departure from the aforementioned scenarios necessitates revisiting the approximations under which such a Markovian QME is derived. For instance, in the presence of (i) baths with structure (a.k.a. `colored' spectra) \cite{IlesSmith2014,Benedetti2014,Groszkowski2023}, correlations \cite{Laine2013}, or finite timescales \cite{Vega2017,Bowen2024} and (ii) retardation effects encountered in distributed quantum networks \cite{Zoller02, Sinha2020a, Sinha2020b}, the default evolution is expected to be non-Markovian. In addition, many-body non-equilibrium phenomena in several biological \cite{Chin2013,Abbasi2023}, chemical \cite{Guerin2012}, and even cosmological \cite{Shandera18,Burrage:2018pyg,Brahma:2020zpk,Zarei21,Colas:2022hlq,Burgess:2022nwu,Colas:2022kfu} systems exhibit non-Markovian dynamics due to inevitable frequency- or time-dependent couplings and/or the presence of non-local spatial correlations. Furthermore, with the advent of practical quantum information processing platforms, where strong interactions and fast time-dependent control have become standard functionalities, describing open quantum dynamics of even a few-qubit systems may fall outside the purview of simple Lindblad-type descriptions \cite{Wei2024}. 
%
\par
Given its pervasiveness, there have been several attempts over the last two decades for understanding, diagnosing, and quantifying non-Markovianity \cite{Rivas2014,Breuer2016}. To this end, several measures of non-Markovianity have been proposed \cite{Wolf08, Rivas10, Luo12, Liu13-2, Lorenzo13, Maniscalco2014,Esposito18}, the most popular being the one proposed in Ref.~\cite{Breuer09} referred to as the ``BLP non-Markovianity,'' which aims to quantify non-Markovianity as the \emph{information backflow} from bath to the system. While most of these measures are platform-agnostic and hence, in principle, universally applicable, their experimental implementation and theoretical computability for practical systems remain severely limited. This is because almost all of them require a full tomographic reconstruction of the state of the system, which scales unfavorably with system size and complexity, even when they permit a more physically-motivated interpretation such as the RHP measure \cite{Rivas2010, Gaikwad2024}. Additionally, BLP-like measures require a complicated estimation of piecewise time-integrated dynamics and an optimization procedure over the state space that grows exponentially in system size. This implies that their calculation requires performing state tomography with sufficiently high bandwidth in order to resolve short-time transient features.
\par
Owing to ongoing efforts aimed at extending quantum optimal control \cite{Koch2016,Fux2021}, quantum sensing \cite{Altherr2021,Riberi2022}, and quantum error correction \cite{Oreshkov2007,Hakoshima2021} in the presence of spatio-temporal correlations, there is a growing need to develop non-Markovian diagnostics that can be scalably and reliably deployed for use in the current NISQ-era quantum systems. In this work, we propose such a measure of non-Markovianity that relies solely on steady-state, spectroscopic measurements making it amenable to implementation in a wide variety of systems. It is worthwhile to note that our viewpoint adopted here is quite complementary to other measures proposed in the past: our aim is to quantify the \emph{absence of Markovianity} by looking for deviations from typical spectra obtained for a strictly Markovian evolution, instead of demanding the presence of a dynamical signature of non-Markovianity as privileged by BLP-like measures. {\color{black} Crucially, avoiding a restrictive definition as a positive proof for non-Markovianity allows the proposed measure to identify non-Markovianity in regimes typically misdiagnosed as Markovian by  measures designed to resolve information backflow.} Inspired by the spectral measure, we develop a time-non-local master equation in the frequency domain that retains full system memory unlike Born-Redfield QME. In this sense, the accuracy of the frequency-domain QME is the same as Nakajima-Zwanzig (NZ) equation in time, derived using projection operators. Unlike NZ description, however, we show how the proposed QME enables a direct calculation of the reduced system spectrum \emph{without} going through the calculation of any two-point correlator in the time domain. 
\par
This paper is organized as follows: In Sec.~\ref{sec:MasterEquationSolutionInFrequencyDomain}, we first derive the frequency-domain master equation and detail the spectrum calculation for an arbitrary quantum system. Next, we motivate how the non-Markovian vs. Markovian evolution manifests in the corresponding frequency-domain propagator and hence the spectrum, and quantify this change by proposing the spectral measure of non-Markovianity. In Sec.~\ref{sec:ExampleJCI}, we apply these constructions to a typical cavity/circuit-QED system and present analytical expressions of the system (qubit) spectra for two cases: thermal (`hot' cavity) and correlated (`squeezed' cavity) bath. 
{\color{black} We also discuss the advantages of the spectral measure, in terms of detecting false negatives in the former case and in terms of its versatility in the latter as compared to the BLP non-Markovianity.} 
In Sec.~\ref{sec:SpatiallySeparatedQubits}, we calculate the proposed measure for a waveguide-QED system using the spectrum of the propagating field modes. We conclude with a summary of our main results and comment on potential extensions of the research presented here in Sec.~\ref{sec:Conclusions}. Additional calculational and technical details are presented in Appendices~\ref{apdx:Superoperator_notation_and_algebra}, \ref{apdx:SystemDynamicsFromBornRedfield} and \ref{apdx:SqueezedBathFullDerivation}.
%
%
\section{Frequency-Domain Master Equation}
\label{sec:MasterEquationSolutionInFrequencyDomain}

We consider a composite quantum system defined on the joint Hilbert space, $\mathcal{H}_S\otimes\mathcal{H}_B$, where $\mathcal{H}_S$ and $\mathcal{H}_B$ span system and bath degrees of freedom, respectively. In the Schr\"odinger picture, the full density matrix, $\chi(t)$, evolves according to the equation of motion,
\begin{IEEEeqnarray}{rCl}
    \partial_t\chi(t)
    &=&
    \qty(\SO{L}_S+\SO{L}_B+\SO{V})\chi(t),
    \label{eq:Sys_Ev_Full}
\end{IEEEeqnarray}
%
where we have defined superoperators [see Appendix \ref{apdx:Superoperator_notation_and_algebra}]
\begin{IEEEeqnarray}{rCl}
    \SO{L}_S
    &\equiv&
    -i\qty[H_S,\> \vdot],
    \startsub \\
    \SO{L}_B
    &\equiv&
    -i\qty[H_B,\> \vdot]+\SO{D}_B ,
    \sub \\
    \SO{V}
    &\equiv&
    -i\qty[V,\> \vdot]. 
    \sub
\end{IEEEeqnarray}
Here $H_S$ and $H_B$ govern the free evolution of the system and bath, respectively, while 
{\color{black}
\begin{IEEEeqnarray}{rCl}
    V
    &=&
    \sum_i O_{S, i}\otimes O_{B, i},
\end{IEEEeqnarray}
}%
denotes the system-bath interaction, all of which are assumed to be time-independent. In addition, we consider a Markovian dissipation on the bath described by the superoperator, $\mathcal{D}_B$, which models some pre-coarse grained bath degrees of freedom. While here we consider a system with strictly unitary evolution, this derivation can be easily adapted to include some local decay, $\SO{D}_S$, on the system too \cite{Thorbeck2024}. We now move into the interaction picture with respect to the free Liouvillian, $\SO{L}_0=\SO{L}_S+\SO{L}_B$, transforming Eq.~(\ref{eq:Sys_Ev_Full}) to 
\begin{IEEEeqnarray}{rCl}
    \partial_t\chi_I(t)
    &=&\SO{V}_I(t)\chi_I(t),
    \label{eq:rho_int}
\end{IEEEeqnarray}
with the interaction-frame density matrix and superoperator defined via the following non-unitary transformations,
\begin{IEEEeqnarray}{rCl}
    \chi_I(t)
    &=&
    e^{-\SO{L}_0t}\chi(t),
    \startsub \\
    \SO{V}_I(t)
    &=&
    e^{-\SO{L}_0t}\SO{V}e^{\SO{L}_0t}.
    \sub
\end{IEEEeqnarray}
When formally integrated and substituted back into Eq.~(\ref{eq:rho_int}), we obtain an integro-differential equation for the joint density matrix,
\begin{IEEEeqnarray}{rCl}
    \partial_t\chi_I(t)
    &=&
    \SO{V}_I(t)\chi_I(0)+\int_0^t\dd{s}\SO{V}_I(t)\SO{V}_I(s)\chi_I(s).
\end{IEEEeqnarray}
As per standard adiabatic elimination methods for deriving the master equation, we now make the `quasi-static bath approximation,' wherein we assume that $\chi(t)$ factorizes at all times, $\chi_I(t)\approx\rho_I(t)\rho_B^{ss}$, and that the bath remains in its steady-state, $\SO{L}_B\rho_B^{ss}=0$. This allows us to perform a partial trace over the bath degrees of freedom leading to the following master equation for the system, 
\begin{IEEEeqnarray}{rCl}
    \partial_t\rho_I(t)
    &=&
    \int_0^t\dd{s}\Tr_B[\>\SO{V}_I(t)\SO{V}_I(s)\rho_I(s)\rho_B^{ss}]. 
    \label{eq:BornIP}
\end{IEEEeqnarray}
Here we have assumed, without loss of generality, that $\Tr[\SO{V}\rho_B^{ss}]=-i\ev{O_B}_{ss}\qty[O_S, \> \vdot]=0$, which effectively normal-orders $O_B$, for interactions linear in $O_B$ considered in this work~\cite{Bowen2024, Boyanovsky:2015xoa}. This `centered bath approximation' ensures that only the integral term remains following the partial trace. Solving Eq.~(\ref{eq:BornIP}) is typically non-trivial due to its time-non-local nature. To alleviate this issue, the substitution $\rho_I(s)\to\rho_I(t)$, is often made, resulting in the Born-Redfield master equation. This is a standard tool for studying non-Markovian effects in the time domain [see Appendix \ref{apdx:SystemDynamicsFromBornRedfield}]. 
\par
In this work, our aim is retain the \emph{full time-non-locality} including the history in the system state. To this end, we develop a frequency-domain master equation as described next. We first transform Eq.~(\ref{eq:BornIP}) back to the lab frame, 
\begin{IEEEeqnarray}{rCl}
    \partial_t\rho(t)
    &=&
    \SO{L}_S\rho(t)
    \nonumber \\
    &&
    +\int_0^t\dd{s}\Tr_B[\>\SO{V}\SO{V}_I(s-t)e^{\SO{L}_0(t-s)}\rho(s)\rho_B^{ss}]. \hspace{8pt}
    \label{eq:RME_DM}
\end{IEEEeqnarray}
For convenience, we then recast our master equation in terms of the vectorized density matrix, $\rho(t)\to\kket{\rho(t)}$, where $\kket{\ldots}$ denotes the column-vectorization of the operator inside [see Appendix~\ref{apdx:Superoperator_notation_and_algebra}]. Expressing Eq.~(\ref{eq:RME_DM}) in this notation, we have
\begin{IEEEeqnarray}{rCl}
    \partial_t\kket{\rho(t)}
    &=&
    \SO{L}^{(0)}\kket{\rho(t)}+\int_0^t\dd{s}\>\SO{K}^{(2)}(t-s)\kket{\rho(s)},\hspace{10pt}
    \label{eq:NLME_Vec}
\end{IEEEeqnarray}
where we have defined the zeroth-order Liouvillian describing free evolution as $\SO{L}^{(0)}\equiv\SO{L}_S$, and the second-order memory kernel describing interaction-induced dynamics as $\SO{K}^{(2)}(t)\equiv\Tr_B\qty[\SO{V}\SO{V}_I(-t)\rho_B^{ss}]e^{\SO{L}_St}$. Due to the non-local form of Eq.~(\ref{eq:NLME_Vec}), the evolution at any point in time is dictated by the entire history of the system. Taking a (unilateral) Fourier transform we obtain a frequency-local equation,
\begin{IEEEeqnarray}{rCl}
    i\omega\kket{\rho[\omega]}-\kket{\rho(0)}
    &=&
    \SO{L}^{(0)}\kket{\rho[\omega]}+\SO{K}^{(2)}[\omega]\kket{\rho[\omega]}. \hspace{14pt}
    \label{eq:FDME}
\end{IEEEeqnarray}
with the transformation and its inverse defined explicitly as
\begin{IEEEeqnarray}{rCl}
    \mathcal{F}_+\qty{A(t)} 
    \equiv 
    A[\omega]
    &=&
    \int_0^\infty\dd{t} e^{-i\omega t}A(t)
    \startsub \\
    \mathcal{F}^{-1}_+\qty{A[\omega]}
    =
    A(t)
    &=&
    \frac{1}{2\pi}\int_{-\infty-i\epsilon}^{\infty-i\epsilon}\dd{\omega}e^{i\omega t}A[\omega].\hspace{16pt}
    \sub
\end{IEEEeqnarray}
Note that we have not set $t\to\infty$ in the upper limit of the integral in Eq.~(\ref{eq:NLME_Vec}), but used the convolution theorem \cite{Shiff},
\begin{IEEEeqnarray}{rCl}
    f[\omega]g[\omega]
    &=&
    \mathcal{F}_+\qty{ \int_0^t\dd{s} f(t-s)g(s)}.
\end{IEEEeqnarray}
Here we distinguish functions in frequency domain by use of square brackets for their arguments. It is important to note that throughout this paper, we use the convention
\begin{IEEEeqnarray}{rCl}
    A^\dag[\omega]
    &=&
    \SO{F}_+\qty{A^\dag(t)} 
    \startsub \\
    A[\omega]^\dag
    &=&
    \qty(\SO{F}_+\qty{A(t)})^\dag=A^\dag[-\omega].
    \sub
\end{IEEEeqnarray}
Eq.~(\ref{eq:FDME}) is a strictly algebraic matrix equation, easily solved for a given initial state,
\begin{IEEEeqnarray}{rCl}
    \kket{\rho[\omega]}
    &\equiv&
    \SO{U}[\omega]\kket{\rho(0)}, 
    \label{eq:FDME_soln}
\end{IEEEeqnarray}
where we define the frequency-domain propagator,
\begin{IEEEeqnarray}{rCl}
    \SO{U}[\omega]
    &\equiv&
    \qty(i\omega\mathbb{I}-\SO{L}^{(0)}-\SO{K}^{(2)}[\omega])^{-1}.
\end{IEEEeqnarray}
{\color{black}This propagator is an open-system analogue to the Hamiltonian resolvent for a time-dependent Hamiltonian and contains explicit information about two-time bath correlators.} Notice that in the frequency domain, density matrices are not positive semi-definite or even Hermitian, and have trace $\Tr[\rho[\omega]]=(i\omega)^{-1}$ [see Eq.~(\ref{eq:HS_Inner_Product})].

Similarly, we write a frequency-domain adjoint master equation describing the evolution of operators in the Heisenberg picture \cite{BreuerPetruccione},
\begin{IEEEeqnarray}{rCl}
    \kket{O_H[\omega]}
    &=&
    \SO{U}^\dag[\omega]\kket{O_S}. 
    \label{eq:FDME_soln_ops}
\end{IEEEeqnarray}
While it is not tractable, in general, to transform Eqs. (\ref{eq:FDME_soln}) and (\ref{eq:FDME_soln_ops}) back to the time domain,  this is not necessarily a limitation for describing non-Markovian evolution. In the next sections, we describe how to leverage this frequency-local description of quantum states to directly calculate and diagnose non-Markovian spectral signatures.

%
%

\subsection{Spectra from correlators in frequency domain}
\label{subsec:SpetraFromCorrelatorsInFrequencyDomain}



The steady state spectrum of a system is calculated by taking the bilateral Fourier transform of the relevant two-time correlator in the steady state $\ev{O^\dag(\tau)O(0)}_{ss}$ \cite{BowenMilburn},
\begin{IEEEeqnarray}{rCl}
    S_{OO}[\omega]
    &=&
    \int_{-\infty}^\infty\dd{\tau}e^{-i\omega\tau}\ev{O^\dag(\tau)O(0)}_{ss}.
    \label{eq:Spec_Gen}
\end{IEEEeqnarray}
In Markovian systems, this is usually a simple task and results in a superposition of Lorentzian distributions. In non-Markovian systems, however, this is typically not the case and {\color{black} it is well know that the quantum regression formula, the standard means for calculating the evolution of multi-time correlators, is invalid} \cite{McCutcheon16}. Even for a simple two-level system undergoing non-Markovian evolution, analytical calculations of these correlators can be quite cumbersome [see Appendix \ref{apdx:SystemDynamicsFromBornRedfield}]. Since Eq.~(\ref{eq:FDME_soln}) already presents a solution for the state in the frequency domain, it allows us to bypass the calculation in time-domain entirely and obtain the spectrum directly.
\par
To this end, we first calculate the steady-state system density matrix from the frequency-domain solution using the final value theorem \cite{Shiff}, 
\begin{IEEEeqnarray}{rCl}
    \kket{\rho_{ss}}
    \equiv 
    \lim_{t\to\infty}\kket{\rho(t)}
    &=&
    \lim_{\omega\to 0}i\omega\kket{\rho[\omega]}.
    \label{eq:FVT}
\end{IEEEeqnarray}
Using this we calculate the correlator,
\begin{IEEEeqnarray}{rCl}
    \SOev{O[-\omega]}{O(0)\rho_{ss}}
    &=&
    \Tr[O[-\omega]^\dag O(0)\rho_{ss}] 
    \nonumber \\
    &=&
    \int_0^\infty\hspace{-6pt} \dd{\tau} e^{-i\omega\tau}\ev{O^\dag(\tau)O(0)}_{ss},\hspace{14pt}
    \label{eq:Spec_TO}
\end{IEEEeqnarray}
where we have defined the adjoint, $\bbra{O}\equiv\kket{O}^\dag$, implying $\SOev{A}{B}=\Tr[A^\dag B]$. Though this resembles Eq.~(\ref{eq:Spec_Gen}), it captures only the time-ordered part. We can calculate the anti-time-ordered part using the identity $\ev{O^\dag(-\tau)O(0)}_{ss}=\ev{O^\dag(\tau)O(0)}_{ss}^*$. From this one can show,
\begin{IEEEeqnarray}{rCl}
    \SOev{O[-\omega]}{O(0)\rho_{ss}}^*
    &=&
%
    \int_{-\infty}^0\dd{\tau} e^{-i\omega\tau}\ev{O^\dag(\tau)O(0)}_{ss}.
    \nonumber \\
    \label{eq:Spec_ATO}
\end{IEEEeqnarray}
Adding Eqs.~(\ref{eq:Spec_TO}) and (\ref{eq:Spec_ATO}), we obtain the final expression for the steady-state spectrum,
\begin{IEEEeqnarray}{rCl}
    S_{OO}[\omega]
    &=&
    \SOev{O[-\omega]}{O(0)\rho_{ss}}+\SOev{O[-\omega]}{O(0)\rho_{ss}}^*
    \nonumber \\
    &=&
    2\Re \bbra{O}\SO{U}[\omega]\kket{O(0)\rho_{ss}}. 
    \label{eq:Spec_Fin}
\end{IEEEeqnarray}
%
%
In Appendix \ref{apdx:SystemDynamicsFromBornRedfield}, we show how the spectrum calculated using Eq.~(\ref{eq:Spec_Fin}) captures the non-Markovian features more accurately than the time-local Born-Redfield calculation. 

%
%

\subsection{Spectral measure of non-Markovianity}
\label{subsec:SpectralMeasureOfNonMarkovianity}

%

It is instructive to compare the form of the frequency-domain propagator $\mathcal{U}[\omega]$ for Markovian vs. non-Markovian evolution. Transforming a Markovian master equation to frequency domain yields a solution similar to Eq.~(\ref{eq:FDME_soln}) but, instead of a frequency-dependent kernel $\SO{K}^{(2)}[\omega]$, it has a constant, second-order Liouvillian,
\begin{IEEEeqnarray}{rCl}
    \SO{U}[\omega]
    &\to&
    \qty(i\omega\mathbb{I}-\SO{L}^{(0)}-\SO{L}^{(2)})^{-1},
\end{IEEEeqnarray}
where $\SO{L}^{(2)}$ is obtained by setting the upper limit of the integral to infinity in Eq.~(\ref{eq:BR_from_Kernel}). 
This form immediately suggests a spectral form given by a sum of Lorentzian peaks, with the respective center frequencies (widths) set by imaginary (real) part of the eigenvalues of the full Liouvillian of the system, $\SO{L}^{(0)}+\SO{L}^{(2)}$. Crucially, the interaction-induced dynamics only lead to constant shifts to widths and positions of the peaks. In the non-Markovian case, however, the additional frequency dependence can qualitatively alter the shape of the spectrum, for instance, imprinting bath spectral features on the system spectrum. 
\par
This suggests that the steady state spectrum in Eq.~(\ref{eq:Spec_Fin}) can provide a simple and robust means to discern the effects of non-Markovian evolution. To define a spectral measure for non-Markovianity, we normalize the spectrum to unit area, representing the probability density of the system emitting an excitation at a given frequency. We then quantify the difference between the Markovian and non-Markovian spectra using the relative entropy, or Kullback-Leibler divergence,
\begin{IEEEeqnarray}{rCl}
    D_{\text{KL}}\qty(S[\omega]||S_{\mathcal{M}}[\omega])
    &=&
    \int_{-\infty}^{\infty}\dd{\omega} S[\omega]\log_2\frac{S[\omega]}{S_{\mathcal{M}}[\omega]},\hspace{14pt}
\end{IEEEeqnarray}
which represents an absolute `distance' between the two spectra \cite{CoverThomas}. 

{\color{black}With an emphasis on an \emph{observable} signature, we take a ratio of this relative entropy to the characteristic bandwidth, $\Omega_{\mathcal{M}}$, given by the spectral gap of the reduced system in the Markovian limit. In a theoretical context, this is obtained by diagonalizing the induced Liouvillian, while in experiment it may be calculated from the FWHM of the emission spectrum.} Doing so privileges sharp, persistent spectral signatures over broad, transient ones. This leads us to the key result of this work: our definition of the spectral measure of non-Markovianity given by
\begin{IEEEeqnarray}{rCl}
    \mathcal{N}_S
    &=&
    \frac{D_{\text{KL}}\qty(S[\omega]||S_{\mathcal{M}}[\omega])}{\Omega_{\mathcal{M}}} 
    \label{eq:SpectralMeasure}.
\end{IEEEeqnarray}
The above measure enjoys a clear, information-theoretic interpretation: it is the information cost per unit bandwidth of making the Markov approximation. Also notice as $S[\omega]\to S_{\mathcal{M}}[\omega]$, we find $\mathcal{N}_{\text{S}}\to 0$ identically. We also stress that this is \emph{not a measure of information backflow}, but rather a measure of distinguishability from Markovian dynamics. While information backflow is indeed a valuable witness, it is not a necessary condition for non-Markovianity \cite{Wolf08, Rivas11, Esposito18}. 
%
%
%

\section{Example: Jaynes-Cummings interaction}
\label{sec:ExampleJCI}

%
We now present an example of constructing the frequency-domain master equation for an archetypal light-matter interaction: a single qubit interacting with a lossy cavity via an interaction Hamiltonian of the form
\begin{IEEEeqnarray}{rCl}
    V
    &=&
     g\qty(\sigma_+a+\sigma_-a^\dag),
\end{IEEEeqnarray}
where $\sigma_+$ and $\sigma_-$ ($a^\dag$ and $a$) are the qubit (cavity) raising and lowering operators. The lossy cavity, with a finite energy relaxation rate, presents an engineered non-Markovian bath to the qubit. Using strong coupling to such ancilla systems, subject to Markovian decay, has been explored earlier to emulate strong non-Markovian effects \cite{Tamascelli2018}.
\par
In the following sections we look at two fundamentally distinct examples of Gaussian ba: (i) a semi-classical bath characterized by a thermal state, which lacks any coherence in the steady-state (diagonal $\rho_B^{ss}$), and (ii) a truly quantum bath characterized by a \emph{squeezed} thermal state which exhibits coherences in the steady-state (non-zero off-diagonal elements in $\rho_B^{ss}$). In both cases, we use $\mathcal{N}_S$ defined in Eq.~(\ref{eq:SpectralMeasure}) to quantify non-Markovianity.

%
%

\subsection{Thermal bath}
\label{subsec:ThermalBath}

\begin{figure}[t!]
    \centering
    \includegraphics[width=\columnwidth]{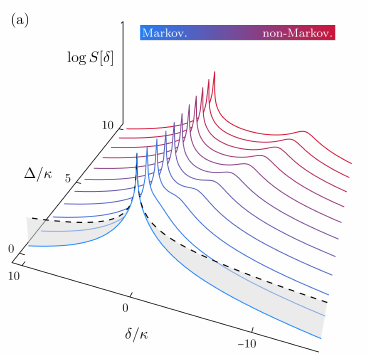} 
    \flushleft
    \includegraphics[width=0.95\columnwidth]{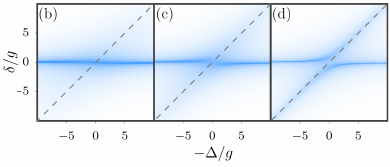}
    
    \caption{
    (a) Steady-state emission spectrum of a qubit interacting with a thermal bath for different values of qubit-cavity detuning. At small $\Delta/\kappa$, the spectrum resembles a single Lorentzian at $\delta=0$. As detuning increases, an asymmetric second order resonance arises near the cavity frequency, becoming more resolved for larger $\Delta/\kappa$. The black, dashed curve shows the Markovian limit [Eq.~(\ref{eq:SpecM})] for zero detuning. Parameters: $\omega_q/g = 2\times 10^5$, $\kappa/g=10$, $\bar{N}=0.1$. (b-d) Density plots of the qubit emission spectra shown for $\kappa/g$ = 5, 2, 0.5. The dashed gray line shows the position of the non-Markovian side peak, $\delta\approx-\Delta$.
    }
    \label{fig:SpectrumWaterfallDetuning}
\end{figure}
\begin{figure*}[t!]
    \centering
    \includegraphics[height=0.5\columnwidth]{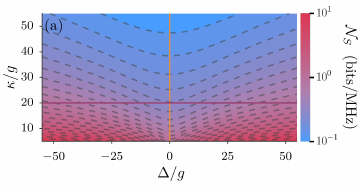}
    \hspace{10pt}
    \includegraphics[height=0.5\columnwidth]{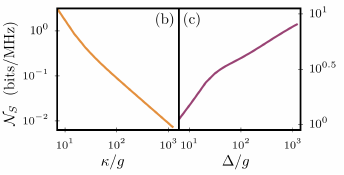}
    \caption{
        (a) Spectral measure of non-Markovianity plotted as a function of qubit-cavity detuning and cavity linewidth. Line cuts showing variation of $\mathcal{N}_{S}$ with (b) bath linewidth $\kappa$, and (c) system-bath detuning $\Delta$. Non-Markovianity decreases as $\mathcal{N}_S\propto(\kappa/g)^{-1}$ for large $\kappa/g$, while it increases as $\mathcal{N}_S\propto \log(\Delta/g)$ for large $\Delta/g$. Parameters: $\omega_q/g=2\times10^5$, $\bar{N}=0.1$. 
        }
    \label{fig:SpectralMeasureThermal}
\end{figure*}
The unitary evolution of the (system) qubit and (engineered bath) cavity are described by  $H_S =  -(\omega_q/2)\sigma_z$  and $H_B =\omega_ca^\dag a$.
To stabilize the cavity in a thermal state, $\ev*{a^\dag a}_{ss}=\bar{N}$ and $\ev*{a^2}_{ss}=0$, we subject it to heating and cooling Lindbladian dissipators,
\begin{IEEEeqnarray}{rCl}
    \SO{D}_B
    &=&
    \kappa\qty(\bar{N}+1)\SO{D}[a]\vdot+\kappa\bar{N}\SO{D}[a^\dag]\vdot,
\end{IEEEeqnarray}
where $\kappa$ is the zero-temperature cavity linewidth. The superoperators acting on the qubit in Eq.~(\ref{eq:NLME_Vec}) can be represented as $4\times4$ non-Hermitian matrices, with free evolution described by 
\begin{IEEEeqnarray}{rCl}
    \SO{L}^{(0)}
    &=&
    \mqty(
        0           &           &           &           \\
                    &i\omega_q  &           &           \\
                    &           &-i\omega_q &           \\
                    &           &           &0
    ),
\end{IEEEeqnarray}
and interaction-induced memory kernel given by
\begin{IEEEeqnarray}{rCl}
    \SO{K}^{(2)}(t)
    &=&
    \smqty(
        \SO{K}_{11}(t)  &                   &                   &-\SO{K}_{44}(t)    \\
                        &\SO{K}_{22}(t)     &                   &                   \\
                        &                   &\SO{K}_{22}^*(t)   &                   \\
        -\SO{K}_{11}(t) &                   &                   &\SO{K}_{44}(t)
    ). 
    \label{eq:Kernel_Therm}
\end{IEEEeqnarray}
Here
\begin{IEEEeqnarray}{rCl}
    \SO{K}_{11}(t)
    &=&
    -2\bar{N} g^2e^{-\kappa t}\cos(\Delta t), 
    \label{eq:Therm_Kernel_Time_Domain}\startsub  \\
    \SO{K}_{44}(t)
    &=&
    -2\qty(\bar{N}+1) g^2e^{-\kappa t}\cos(\Delta t), 
    \sub  \\
    \SO{K}_{22}(t)
    &=&
    -\qty(2\bar{N}+1) g^2e^{-(\kappa -i\omega_c)t},
    \sub
\end{IEEEeqnarray}
with $\Delta\equiv\omega_q-\omega_c$ denoting the qubit-cavity detuning. 
Here the $\SO{K}_{11}(t)$ and $\SO{K}_{44}(t)$ terms correspond to the heating and cooling rates, respectively, while the imaginary part of $\SO{K}_{22}(t)$ corresponds to the effective shift of the qubit frequency [see Eq.~(\ref{eq:BR2ndOrderLiouvillian2})].
\par
Taking the Fourier transform and expressing in terms of the detuning from the bare qubit frequency, $\delta=\omega-\omega_q$, we obtain,
\begin{IEEEeqnarray}{rCl}
    \SO{K}_{11}[\delta]
    &=&
    -\frac{2\bar{N} g^2\qty(\kappa+i(\delta+\omega_q))}{\qty(\kappa+i(\delta+\omega_q))^2+\Delta^2}, 
    \startsub  \\
    \SO{K}_{44}[\delta]
    &=&
    -\frac{2\qty(\bar{N}+1) g^2\qty(\kappa+i(\delta+\omega_q))}{\qty(\kappa+i(\delta+\omega_q))^2+\Delta^2}, 
    \sub  \\
    \SO{K}_{22}[\delta]
    &=&
    -\frac{\qty(2\bar{N}+1) g^2}{\kappa+i\qty(\delta+\Delta)}. 
    \sub
\end{IEEEeqnarray}
%

Following the procedure in Sec. \ref{subsec:SpetraFromCorrelatorsInFrequencyDomain}, we use Eq.~(\ref{eq:FVT}) to find, for an arbitrary $\kket{\rho(0)}$, the steady-state density matrix,
\begin{IEEEeqnarray}{rCl}
    \kket{\rho_{ss}}
    &=&
    \mqty(
        \frac{\bar{N}+1}{2\bar{N}+1}    &0                          
        &
        0                               &\frac{\bar{N}}{2\bar{N}+1}
    )^\T,
    \label{eq:Thermal_SS_DM}
\end{IEEEeqnarray}
which is exactly as that expected for a qubit coupled to a thermal bath. Now applying Eq.~(\ref{eq:Spec_Fin}) with $O=\sigma_-$, the normalized steady-state qubit emission spectrum can be written in terms of matrix elements of the kernel as,
\begin{IEEEeqnarray}{rCl}
    S[\delta]
    &=&
    \frac{1}{\pi}
    \frac{
        -\Re\SO{K}_{22}[\delta]
    }{
        \qty(\delta-\Im\SO{K}_{22}[\delta])^2+\qty(\Re\SO{K}_{22}[\delta])^2
    }.
    \label{eq:FD_Therm_Spec_Kernel}
\end{IEEEeqnarray}
Interestingly, the overall form of the spectrum can be thought of as a `nested' Lorentzian where the width ($\Re\SO{K}_{22}[\delta]$) and the center frequency ($\Im\SO{K}_{22}[\delta]$) themselves are Lorentzian and Fano functions of the qubit detuning, respectively. A plot of the resulting spectral profile shown in Fig. \ref{fig:SpectrumWaterfallDetuning}, shows a second order resonance near the cavity frequency, $\delta\approx-\Delta$, with width given by the cavity decay rate, $\kappa$. This also indicates that while non-Markovianity can be considered as a time-dependent modulation of the system response, which renders the system non-static even when analyzed in its own interaction frame, this is not a simple sideband modulation of system frequency at system-bath detuning since that would have created an image of bare system resonance with width $\gamma_{\rm eff}$ at $\delta\approx-\Delta$. Instead this resonance corresponds to the emission spectrum of the cavity observed through that of the system; in fact, the isolated bath correlator, $\ev{a^\dag[\delta]a}=\SO{K}_{22}[\delta]$. As shown in Figs.~\ref{fig:SpectrumWaterfallDetuning}(b-d), the non-Markovian side peak tracks the normal mode splitting at high coupling strengths, highlighting that the origin of non-Markovianity in canonical QED setups is the coherent system-bath coupling with its magnitude scaling as effective coupling strength, $g/\sqrt{\Delta^{2} + \kappa^{2}}$.
\par
We can gain some insight by comparing the form of Eq.~(\ref{eq:FD_Therm_Spec_Kernel}) to the well known Markovian spectrum, $S_{\mathcal{M}}[\delta]$ obtained from the Born-Markov master equation,
\begin{IEEEeqnarray}{rCl}
    S_{\mathcal{M}}[\delta]
    &=&
    \frac{1}{\pi}\frac{\gamma_{\text{eff}}}{\qty(\delta-\delta_{\text{eff}})^2+\gamma_{\text{eff}}^2}, 
    \label{eq:SpecM}
\end{IEEEeqnarray}
with Lamb shift and induced decay rate given by
\begin{IEEEeqnarray}{rCl}
    \delta_{\text{eff}}
    &=&
    \Im\SO{K}_{22}[0]
    =
    \frac{(2\bar{N}+1) g^2\Delta}{\Delta^2+\kappa^2}
    \startsub, \\
    \gamma_{\text{eff}}
    &=&
    -\Re\SO{K}_{22}[0]
    =
    \frac{(2\bar{N}+1) g^2\kappa}{\Delta^2+\kappa^2},
    \sub 
\end{IEEEeqnarray}
respectively. Interestingly, we see that the Markovian limit may be recovered by simply evaluating the kernel at the qubit frequency in Eq.~(\ref{eq:FD_Therm_Spec_Kernel}). 



We also note that the asymptotic behavior is different from the Markovian case. For the Lorentzian profile, $S_{\mathcal{M}}[\delta]\propto\delta^{-2}$ in the limit of $\delta\to\pm\infty$, however in the non-Markovian spectrum we find $S[\delta]\propto\delta^{-4}$. Since the two spectra are normalized to unit-area, the different end behavior accounts for the extra area beneath the side peak at $\delta = -\Delta$ in Fig.~\ref{fig:SpectrumWaterfallDetuning}.

It is already clear from the qualitative behavior of the spectra shown in Fig. \ref{fig:SpectrumWaterfallDetuning} that the distinguishing spectral feature of non-Markovian evolution is an additional peak appearing near the cavity frequency. As $\Delta/\kappa$ decreases the side peak becomes less distinguishable from the central Lorentzian described by Eq.~(\ref{eq:SpecM}), leading to suppressed non-Markovianity. We now apply our spectral measure defined in Eq.~(\ref{eq:SpectralMeasure}) to provide a quantitative estimate of this change in non-Markovianity. Fig.~\ref{fig:SpectralMeasureThermal}(a) summarizes our results showing an increase in $\mathcal{N}_{s}$ with decreasing cavity linewidth and increased qubit-cavity detuning.  Given that the Markov approximation is justified for a fast decay of bath correlators, a rapid decrease of $\mathcal{N}_S$ towards zero for increasing linewidth is consistent with this expectation. Fig.~\ref{fig:SpectralMeasureThermal}(b) shows this monotonic decrease on a log scale with $\mathcal{N}_S\propto\kappa^{-1}$. 

Fig. \ref{fig:SpectralMeasureThermal}(c) shows a monotonic increase in $\mathcal{N}_S$ with qubit-cavity detuning since the side peak is more distinct for higher detunings. However, this increase with system-bath detuning slows at large detunings, since as the side peak moves farther away from zero (qubit) frequency, it also shrinks in height. 

It is worthwhile to note that the calculation of $\mathcal{N}_S$ is not limited to using the frequency-domain master equation. In fact, it can be evaluated for non-Markovian spectra obtained using any means such as the one obtained using Born-Redfield master equation. However, as shown in Appendix \ref{apdx:SystemDynamicsFromBornRedfield}, not only is the BR-QME route of calculating the spectra via the Fourier transform of the two-time system correlator more cumbersome, it systematically shows a small error in predicting the position of the non-Markovian side peak [see Fig.~\ref{fig:Compare_Spectra}].

%
%

\subsection{Squeezed bath}
\label{subsec:SqueezedBath}

%
\begin{figure*}[t!]
    \centering
    \includegraphics[width=\columnwidth]{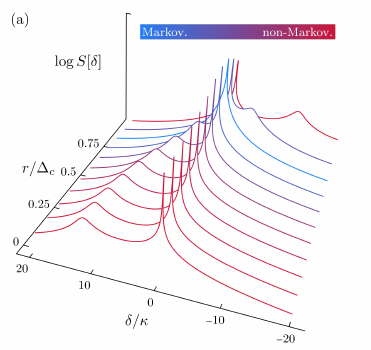}
    \includegraphics[width=\columnwidth]{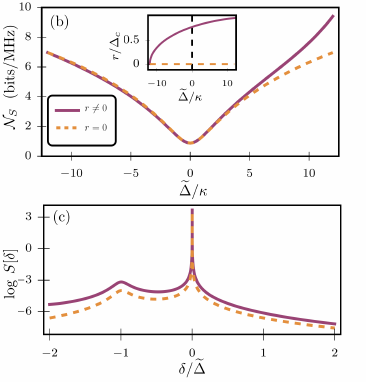}
    \caption{
        (a) Steady-state emission spectrum of a qubit interacting with a squeezed cavity for different values of the squeezing parameter. As $r$ increases, the side peak moves with the effective detuning, $\widetilde\Delta$, leading to a squeezing-induced resonance. This is evidenced by the single peak of the blue curve at $r/\Delta_c\approx0.78$. (b) Spectral measure of non-Markovianity as a function of $\widetilde{\Delta}$. This is done in two ways: (solid, purple) increasing the squeezing parameter while maintaining a constant bare detuning, and (dashed, orange) varying the bare detuning while maintaining zero squeezing. The inset shows the corresponding values of $r$ for the given values of effective detuning, $\~\Delta$. (c) The spectra resulting from these two different means of varying the effective detuning with $\widetilde{\Delta}/\kappa=10$. Parameters: $\Delta_q/g=200$, $\Delta_c/g=320$, $\kappa/g=10$. 
    }
    \label{fig:SpectrumWaterfallSqueezing}
\end{figure*}%

As a second example, we study the same two-level system interacting with a more general bath. Rather than a purely diagonal bath state, we introduce quantum correlations in the steady state of the bath while still maintaining its Gaussianity by introducing a two-photon drive on the cavity with pump frequency, $\omega_p\approx2\omega_c$. Moving to a frame rotating at $\omega_p/2$, the Hamiltonians take the form \cite{Stern22}
\begin{IEEEeqnarray}{rCl}
    \label{eq:Squeeze_Hamiltonian}
    H_S
    &=&
    -\frac{\Delta_q}{2}\sigma_z,
    \startsub \\
    H_B
    &=&
    \Delta_c a^\dag a+\frac{r}{2}\qty(a^2+a^{\dag2}),
    \sub
\end{IEEEeqnarray}
with detunings $\Delta_i=\omega_i-\omega_p/2$. This particular form is chosen because it is applicable in several architectures, namely SQUID-based superconducting circuit resonators, where the squeezing strength, $r$, can even be made \emph{in-situ} tunable. Without loss of generality, here we assume $r \in \mathbb{R}^+$ with $r<|\Delta_c|$. For mathematical simplicity, further we perform a Bogoliubov transformation to diagonalize the bath Hamiltonian [Eq.~(\ref{eq:Bogo})],
\begin{IEEEeqnarray}{rCl}
    H_B
    &=&
    \widetilde{\Delta}_c\~a^\dag\~a.
\end{IEEEeqnarray}
Consequently, the interaction Hamiltonian transforms to,
\begin{IEEEeqnarray}{rCl}
    V
    &=&
     g_1(\sigma_+\~a+\sigma_-\~a^\dag)+ g_2(\sigma_+\~a^\dag+\sigma_-\~a),
    \label{eq:JCIBogoliubov} 
\end{IEEEeqnarray}
with modified detuning and coupling strengths,
\begin{IEEEeqnarray}{rCl}
    \widetilde{\Delta}_c
    &=&
    \sqrt{\Delta_c^2-r^2}
    \label{eq:Delta_C_Eff} \startsub, \\
     g_1
    &=&
    g \sqrt{\frac{\Delta_c}{2\widetilde{\Delta}_c} +\frac{1}{2}}
    \sub, \\
     g_2
    &=&
     g \sqrt{\frac{\Delta_c}{2\widetilde{\Delta}_c} -\frac{1}{2}}.
    \sub
\end{IEEEeqnarray}
Even with only a cooling dissipator on the cavity, $\SO{D}_B=\kappa\SO{D}[a]\vdot$, the two-photon drive ensures non-zero steady-state population in the cavity which is essentially the noise photons associated with a squeezed thermal state. The corresponding dissipator in the Bogoliubov frame can be written as,
\begin{IEEEeqnarray}{rCl}
    \SO{D}_B
    &=&
    \kappa\qty(\qty(\bar{N}+1)\SO{D}[\~a]\vdot+\bar{N}\SO{D}[\~a^\dag]\vdot\>)
    \nonumber \\
    &&
    -\qty(\kappa+i\widetilde{\Delta}_c)\bar{M}\qty(\SO{S}[\~a]\vdot+\SO{S}[\~a^\dag]\vdot\>),
    \label{eq:Sq_Env_Dissipator}
\end{IEEEeqnarray}
{\color{black} where have defined the steady-state bath excitations number $\langle\tilde{a}^\dag \tilde{a}\rangle_{ss}=\bar{N}$, and steady-state bath coherence $\langle\tilde{a}^2\rangle_{ss}=\bar{M}$ [see Appendix~C].} In addition to the conventional heating and cooling Lindbladian dissipators, we have two additional dissipators of the form $\SO{S}[O]\vdot=2O\vdot O-O^2\vdot-\vdot O^2$, which we call the `squeezing' dissipators. Rather than directly damping excitations or coherences, these couple the off-diagonal elements of the density matrix. 
\par
Having already studied the effects of cavity decay rate and detuning in the previous section, here we focus on the effect of squeezing-induced bath correlations on system non-Markovianity. Following the same procedure as in the previous section, we obtain the steady-state qubit emission spectrum. As in Eq.~(\ref{eq:FD_Therm_Spec_Kernel}), the spectral form is that of a nested Lorentzian except with a modified form of $\SO{K}_{22}[\delta]$,
\begin{IEEEeqnarray}{rCl}
    \SO{K}_{22}[\delta]
    &=&
    \frac{2 g_1 g_2\bar{M}+ g_1^2(2\bar{N}+1)}{\kappa+i\qty(\delta+\widetilde{\Delta})}
    \nonumber \\
    &&
    +
    \frac{2 g_1 g_2\bar{M}^*+ g_2^2(2\bar{N}+1)}{\kappa+i\qty(\delta+\widetilde\Sigma)},
\end{IEEEeqnarray}
now with $\delta=\omega-\Delta_q$. Note that squeezing leads to a generally complex, non-zero $\bar{M}$ and a tunable $\widetilde{\Delta}=\Delta_q-\~\Delta_c$. This, in turn, corresponds to two Lorentzians and two Fano resonances (proportional to the $\Im[\bar{M}]$) near the difference and sum frequencies of the qubit and the cavity respectively. In addition, the bath correlations lead to non-zero $\bar{\SO{K}}_{32}[\delta]$, which couple the evolution of the off-diagonal elements $\rho_{ge}$, $\rho_{eg}$ of the qubit density matrix. While these lead to a much more complicated expression for the spectrum as shown by the detailed derivation presented in Appendix \ref{apdx:SqueezedBathFullDerivation}, the effect of these terms in the spectrum appears at $\mathcal{O}(g^{4})$ and is, therefore, muted.

Unsurprisingly, the form of the spectral profile in Fig.~\ref{fig:SpectrumWaterfallSqueezing}(a) looks very similar to that obtained for the thermal case since the spectral weights of the additional contributions at the sum frequency are damped due to higher detuning from the central Lorentzian. There is a crucial difference though: Eq.~(\ref{eq:Delta_C_Eff}) tells us that the squeezing strength $r$ determines the effective bath frequency, thus varying $r$ allows changing the position of the side peak. The squeezing parameter also determines the effective coupling strength and bath correlators, but neither have as striking an effect on the spectrum as the squeezing-dependent effective system-bath detuning. From this we expect similar effect on non-Markovianity, whether we vary the bare detuning, $\Delta$, or $\widetilde{\Delta}_c$ by tuning $r$. 

This is confirmed by the spectral measure calculated for these two means of varying $\widetilde{\Delta}$, as shown in Fig. \ref{fig:SpectrumWaterfallSqueezing}(b). For large squeezing strengths, near $\widetilde{\Delta}/\kappa=10$, we do find deviations between the two cases. This is due to effective couplings increasing exponentially with squeezing strength, which should lead to enhanced non-Markovianity: this, in fact, manifests in the increasing spectral weight of the side peak. A key result of this study is that both curves have a minimum when $\widetilde{\Delta}=0$, indicating that memory effects could be diminished by tuning \emph{bath coherences only}. We note, however, that $\widetilde{\Delta}$ increases monotonically with $r$, so this squeezing-induced resonance can only occur if originally the qubit is parked below the cavity, i.e. $\Delta_c>\Delta_q$.

Further, it is well known that the Born-Redfield QME does not, in general, preserve positivity of the system density matrix for all times \cite{Gaspard:1999}. This violation of positivity occurs in the presence of squeezed correlations for certain choice of parameters and initial state, when using the Born-Redfield QME [Eq.~(\ref{eq:BR_Sq_2nd_Order_Dissipators})] derived in Appendix \ref{apdx:SqueezedBathFullDerivation}. Remarkably, the frequency-domain description, on the other hand, preserves positivity at all times for the same choice of parameters (see Fig.~\ref{fig:SqueezedPositivity}). While a full investigation of this issue is beyond the scope of the current work, our results indicate the role played by the Markov approximation in positivity violation of the reduced system state.
%
%
%
\subsection{Comparison to BLP non-Markovianity}
\label{sec:Comparison_to_BLP}
%
In this section, we compare the spectral measure $\mathcal{N}_{s}$ proposed in this work against a time-domain measure for non-Markovian dynamics. For the latter, perhaps the most widely regarded measure for non-Markovianity is the trace distance-based BLP measure \cite{Breuer09, Breuer10}, which relates the resurgence of information to the non-contractivity of the dynamical map, $\SO{U}(t)$, in our language. If a map is a quantum Markov semigroup then it must be purely contractive, and thus any two initial states will become less distinguishable at all times during their evolution. The BLP non-Markovianity quantifies this using the trace distance, 
\begin{IEEEeqnarray}{rCl}
    D_{\Tr}(\rho_1, \rho_2)
    &=&
    \frac12\Tr\qty|\rho_1-\rho_2|,
\end{IEEEeqnarray}
where $\qty|A| =\sqrt{A^\dag A}$. Defining $\rho_{1, 2}(t)=\SO{U}(t)\rho_{1,2}(0)$, for two different initial states, one will find
\begin{IEEEeqnarray}{rCl}
    \dv{t} D_{\Tr}(\rho_1(t), \rho_2(t))
    &\leq&
    0,
\end{IEEEeqnarray}
for a Markovian process. The degree to which this condition is broken can then be used to quantify the non-contractivity, and thus the BLP measure of non-Markovianity is defined as
\begin{IEEEeqnarray}{rCl}
    \SO{N}_{\text{BLP}}
    &=&
    \max_{\rho_{1, 2}(0)}\int_{>0}\dd{t}\dv{t} D_{\Tr}(\rho_1(t), \rho_2(t)).
    \label{eq:BLP}
\end{IEEEeqnarray}
Here the region of integration comprises all times in which the integrand is positive, that is, whenever the trace distance is increasing. A maximization is then performed over all combinations of initial states, both pure and mixed. While extremely generic and independent of any specific representation of open system dynamics, there are some critical disadvantages in applying this measure. Given the nontrivial nature of the integral, typically its calculation is analytically intractable and must be performed numerically. Similarly, the maximization over arbitrary initial states is cumbersome and computationally expensive especially when studying large systems since the state space grows as $\sim d^4$, where $d=\dim(\mathcal{H}_S)$. Moreover, the measure is explicitly tomographic in nature, making it impractical for implementation in experiments. 

\begin{figure}[t!]
    \centering
    \includegraphics[width=0.9\columnwidth]{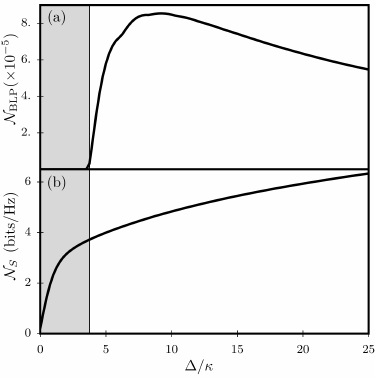}

    \caption{
        (a) BLP non-Markovianity for a qubit interacting with a thermal bath as a function of qubit-cavity. Qubit dynamics are calculated from the Born-Redfield master equation shown in Appendix \ref{apdx:SystemDynamicsFromBornRedfield}. (b) Spectral measure of non-Markovianity as a function of qubit-cavity detuning.  Parameters: $\omega_q/g=2\times10^5$, $\kappa/g=20$, $\bar{N}=0.1$.
        }
    \label{fig:BLPThermal}
\end{figure}
\begin{figure}[t!]
    \centering
    \includegraphics[width=\columnwidth]{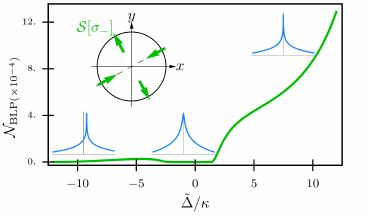}

    \caption{
        The integral from the BLP non-Markovianity with initial states given by the $\sigma_x$ eigenstates evolved according to the Born-Markov master equation. Insets show the corresponding spectra calculated in each of the three regimes, each exhibiting only a central Lorentzian and lacking any non-Markovian features. Parameters: $\Delta_q/g=200$, $\Delta_c/g=320$, $\kappa/g=10$.
        }
    \label{fig:BLPSqueezed}
\end{figure}

We now explicitly compare $\mathcal{N}_{\rm BLP}$ with $\mathcal{N}_{S}$ for the Jaynes-Cummings interaction for the two cases of thermal and squeezed cavity studied in this section. We calculate $\mathcal{N}_{\rm BLP}$ from the time-domain solution to the Born-Redfield master equation, Eq.~(\ref{eq:BR_ME_Thermal}). We perform an abridged calculation by skipping the maximization and simply using the $\sigma_z$ eigenstates, well known to maximize the integral in Eq.~(\ref{eq:BLP}) \cite{Breuer09, Breuer10}, and compare it to $\mathcal{N}_{S}$ calculated using the spectra in Eqs.~(\ref{eq:FD_Therm_Spec_Kernel}) and (\ref{eq:SpecM}) \footnote{As shown by the comparison of the spectra in Fig.~\ref{fig:Compare_Spectra}, using Born-Redfield QME to estimate $\mathcal{N}_{S}$ instead would lead to a slightly lower value but the qualitative trend as a function of detuning is expected to remain the same.}. A direct comparison of magnitudes of the two measures is not instructive since they are fundamentally different quantities; however, we plot them on a logarithmic scale to compare the qualitative behavior as a function of $\Delta/\kappa$ in Fig. \ref{fig:BLPThermal}. As evident, $\mathcal{N}_{\rm BLP}$ shows a non-monotonic behavior, reaching a maximum value at $\Delta \approx 8.5\kappa$; on the other hand, $\mathcal{N}_{S}$ shows a monotonic increase, which is a result of the emphasis on \emph{distinguishability} in its definition. For small $\Delta/\kappa$, both the metrics predict a low value of non-Markovianity; it is worth noting, however, that $\mathcal{N}_{\rm BLP}$ is identically zero in this regime whereas, by definition, the spectral measure cannot vanish completely unless the non-Markovian and Markovian spectra are identical. This suggests that the regime of small detuning, $\Delta/\kappa <$ ``few", represents \emph{non-Markovianity without information backflow} detectable using the spectral measure proposed here.

\begin{figure*}[t!]
    \centering
    \includegraphics[width=\columnwidth]{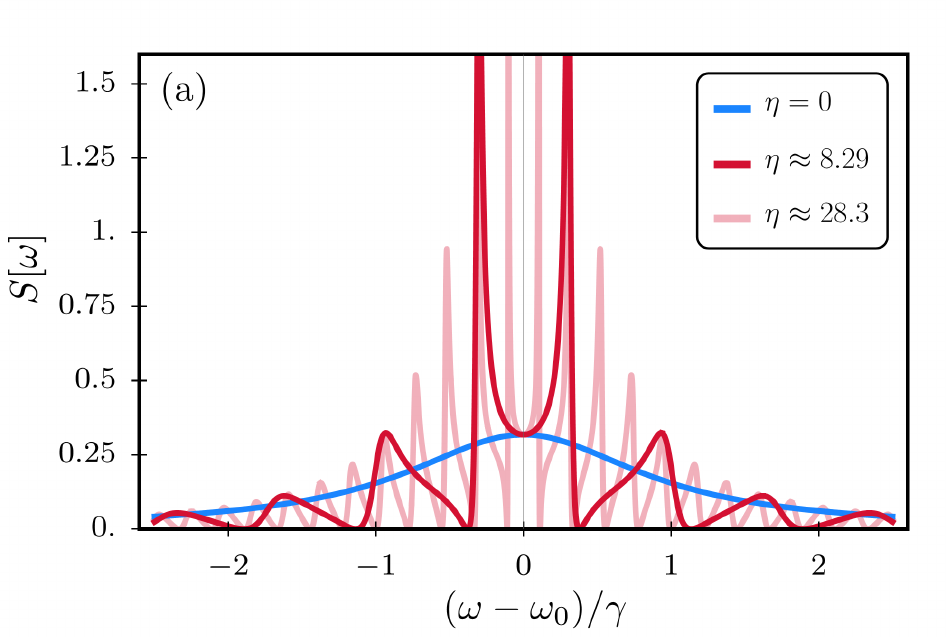}
    \includegraphics[width=\columnwidth]{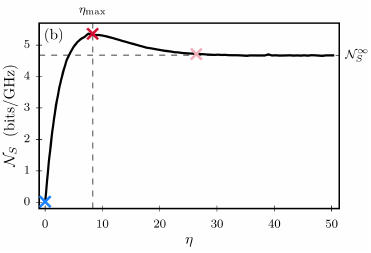}

    \caption{
        (a) Steady-state spectrum of photons emitted into the waveguide by the qubits initialized in the symmetric state for different separations parameterized as non-Markovianity parameter $\eta$: $\eta=0$ corresponds to zero separation and shows a Lorentzian of width $2\gamma$ (blue), while $\eta=28.3$ corresponds to a large separation and shows repeating Fano-like peaks (red). The intermediate regime, for $\eta = 8.29$, corresponds to a large separation and shows Fano resonances which repeat several times within the width of the Lorentzian (light red). (b) Spectral measure of non-Markovianity as as function of qubit separation. The metric maximizes at $\eta=\eta_{\text{max}}$ then decreases and saturates at $\mathcal{N}_S^\infty$. The separation is always chosen such that $\eta\omega_0/\gamma=2\pi n,\>n\in \mathbb{Z}^+_0$. Parameters: $\omega_0/\gamma=500$, $\beta=0.95$.
    }
    \label{fig:SpectralMeasureDelayedFeedback}
\end{figure*}

{\color{black}
    Another subtlety inherent to the BLP non-Markovianity is the assumption  that the map, $\mathcal{U}(t)$, be completely positive and trace preserving, and any map which fails to preserve positivity is guaranteed to result in a non-zero $\mathcal{N}_{\rm BLP}$. Since the BR-QME, the most widely studied non-Markovian master equation, is known to violate positivity in general, $\mathcal{N}_{\rm BLP}$ is not guaranteed to be applicable for all systems. Specifically, in the case of the squeezed cavity, we find that both the BM-QME and the BR-QME fail to preserve positivity for sufficient squeezing strengths [see Appendix~C], and thus $\mathcal{N}_{\rm BLP}$ is guaranteed to be non-zero. 

    To demonstrate this, we calculate an abridged version of the BLP non-Markovianity, forgoing the maximization and simply using initial states given by $|\sigma_x=\pm 1\rangle$. The fully time-independent BR-QME has no history-dependence by definition, but fails to be CPTP for sufficient squeezing. As shown in Fig.~5, this lack of positivity leads to a non-zero $\mathcal{N}_{\rm BLP}$, as expected, while $\mathcal{N}_S$ remains zero by definition. 
}

%
%

\section{Example: Spatially-separated qubits}
\label{sec:SpatiallySeparatedQubits}


Non-Markovianity also manifests in systems with some form of delayed feedback \cite{Vitali99, Zoller02}. That is, systems which interact with some traced out degree of freedom characterized by a finite delay time, $\tau_{\text{B}}$. Such systems display their non-Markovianity clearly, with the evolution of the system at time $t$ depending directly on the state at time $t-\tau_{\text{B}}$.

We look at an example of this type, namely two spatially-separated qubits coupled to a waveguide. The system Hamiltonian is simply given by the sum of the free Hamiltonians of the two qubits, 
\begin{IEEEeqnarray}{rCl}
    H_{S}
    &=&
    \sum_{m=1, 2}\frac{\omega_0}{2}\sigma^z_m.
\end{IEEEeqnarray}
The Hamiltonian for the field modes within the waveguide is given by
\begin{IEEEeqnarray}{rCl}
    H_{B}
    &=&
    \int_0^\infty\dd{\omega}\qty(a^\dag[\omega] a[\omega]+b^\dag[\omega] b[\omega]),
\end{IEEEeqnarray}
where $a[\omega]$ ($b[\omega]$) is the lowering operator of the right (left)-propagating field. The interaction Hamiltonian, expressed in the interaction picture, is given by
\begin{IEEEeqnarray}{rCl}
    V_I(t)
    &=&
    \sum_{m=1, 2}\int_0^\infty\dd{\omega}\Big( g[\omega]e^{i(\omega_0-\omega)t}\sigma_m^+\Big(e^{i\omega x_m/v}a[\omega]
    \nonumber \\
    &&
    \hspace{66pt} +e^{-i\omega x_m/v}b[\omega]\Big) + \text{H.c.}\Big),
\end{IEEEeqnarray}
which shows how the interaction explicitly depends on the positions of the qubits, $x_m$. Here $\sigma_m^+$ is the raising operator for the $m$th qubit. Since field modes propagate with a finite velocity, $v$, in the waveguide, this leads to a delay, $\tau_{\text{B}}\equiv|x_1-x_2|/v$, in excitation exchange between the two qubits. We refer the reader to Ref.~\cite{Sinha2020a} for a detailed solution of time-dependent qubit populations and field amplitudes in the presence of such retardation effects. The expression for the latter in the frequency domain is given by,
\begin{IEEEeqnarray}{rCl}
    c_{a, b}[\omega]
    &=&
    \frac{
        \sqrt{\frac{\gamma\beta}{2\pi}}\cos(\frac{\eta\omega}{2\gamma})
        }{
        \omega-\omega_0-
        \frac{\gamma\beta}{2}\sin(\frac{\eta\omega}{\gamma})+
        i\frac{\gamma}{2}\qty(1\hspace{-2pt}+\hspace{-2pt}\beta\cos(\frac{\eta\omega}{\gamma}))
        },
        \nonumber \\
\end{IEEEeqnarray}
where $\gamma$ denotes the qubit decay rate and $\beta$ represents the waveguide coupling efficiency. As discussed in Ref.~\cite{Sinha2020a}, different regimes of non-Markovianity can be characterized using the parameter $\eta=\gamma\tau_{B}$. Here, we will deploy the spectral measure, $\mathcal{N}_S$, introduced in our work to quantify the non-Markovianity as a function of $\eta$.

Assuming a symmetric initial state, $|c_a[\omega]|^{2}$ represents the joint probability distribution describing the likelihood of detecting a photon of frequency $\omega$ emitted in the right propagating mode, $P(\omega, a)$. The spectrum we are interested in is the conditional probability distribution, $S[\omega]\equiv P(\omega|a)$, corresponding to the probability of a photon emitted at frequency $\omega$. We obtain this by dividing $|c_a[\omega]|^{2}$ by the marginal probability, $P(a)$, 
\begin{IEEEeqnarray}{rCl}
    S[\omega]
    &=&
    \frac{|c_a[\omega]|^2}{\int \dd{\omega'}|c_a[\omega']|^2}.
    \label{eq:Swaveguide}
\end{IEEEeqnarray}
Note that the probability distributions $P(\omega,a),\,P(a)$ do not sum to unity due to finite probability associated with emission into left-propagating field modes and ${\beta < 1}$. Nonetheless, in accordance with the properties of a conditional distribution, $S[\omega]$ is a normalized spectrum as desired per the prescription presented in Sec.~\ref{subsec:SpectralMeasureOfNonMarkovianity}.

In order to calculate $\mathcal{N}_S$, we first calculate the Markovian spectrum $S_{\mathcal{M}}[\omega]$ given by $\eta=0$ and corresponding to no delay in the qubit-qubit interaction. In this limit, the spectrum is a single Lorentzian of width $2\gamma$. As shown in Fig.~\ref{fig:SpectralMeasureDelayedFeedback}(a), with increase in the separation between the qubits, extra resonances appear in the form of repeating Fano line shapes that become more frequent with larger separation. {\color{black} These Fano resonances are a direct result the finite propagation-time, causing interference between photons emitted by the system at different times.}

The resultant $\mathcal{N}_{S}$ as a function of $\eta$ is shown in Fig. \ref{fig:SpectralMeasureDelayedFeedback}(b).
Near $\eta=0$, corresponding to small separation between the atoms, the cooperative decay enhances the central peak width leading to small $\mathcal{N}_{S}$. More interestingly, while increase in $\eta$ increases $\mathcal{N}_{S}$, we see that $\mathcal{N}_S$ maximizes at some $\eta_{\text{max}}$, and then saturates to a lower value, $\mathcal{N}_S^\infty$. Crucially, $\eta_{\rm max}$ corresponds to the situation where the position of the second Fano resonance coincides with the width of the Markovian Lorentzian, $\gamma$. This leads to drastic changes within the central lobe of the spectrum, leading to strong deviations from $S_{\mathcal{M}}$ and largest value of $\mathcal{N}_{S}$. On the other hand, for very large $\eta$, the spectrum becomes a dense frequency comb whose average deviation per unit bandwidth from the Markovian spectrum saturates to a constant value. It is worthwhile to note that the proposed spectral measure is able to demarcate the crossover from the regime of monotonically-increasing non-Markovianity ($\eta \ll \eta_{\rm max}$) to large constant non-Markovianity ($\eta \gg \eta_{\rm max}$).
%
%
%
%
\section{Conclusions}
\label{sec:Conclusions}
In summary, we have proposed a measure for non-Markovianity relying solely on spectroscopic measurements that allows quantifying non-Markovian effects in open quantum systems at late times. We also present a second-order frequency-local master equation that circumvents the need for making any Markov approximation and enables direct calculation of the steady-state system spectrum. The spectral measure of non-Markovianity provides several insights both about the rates entering master equations as well as the regimes of pronounced non-Markovianity in all the examples discussed here. For instance, in the presence of a squeezed cavity, it is able to delineate the regime where the system can exhibit correlation-induced Markovianity for specific values of squeezing strength, while in waveguide-QED systems it exhibits a non-monotonic behavior with emitter separation showing that non-Markovianity is not necessarily maximized by increasing the time-delay due to finite-velocity propagation. {\color{black} These two examples also highlight the versatility of the spectral measure, revealing non-Markovian spectral features on both UV and IR frequency scales. These features correspond to transient short-time effects and persistent long-time effects, respectively.}
\par
Crucially, using the proposed measure we identify regimes where measures privileging recurrences in system dynamics fail to diagnose deviations from Markovianity.
A more stark failure of time-domain detection manifests in the case of a qubit subject to squeezing-type Markovian dissipation where the dynamical oscillation of Bloch vector length is misdiagnosed as non-Markovianity by dynamical measures. 
Spectroscopic measurements are also more appealing for adoption in practical quantum devices \cite{Groblacher2015,Carmele2019}, as they eliminate the need for fast, time-resolved measurements required to capture small transient features, and do not incur significant hardware overhead for extensions to multipartite systems unlike other tomographic measures of non-Markovianity \cite{Breuer09, Wolf08, Rivas10, Luo12, Liu13-2, Lorenzo13, Esposito18}. A natural and compelling application of frequency-domain correlation diagnosis presented here can be in quantum many-body \cite{Quan2006,DallaTorre2010} and field-theoretic \cite{Lombardo:1995fg,Koksma:2009wa,Koksma:2011dy,Lawrence18,Boyanovsky:2015xoa,Agon:2017oia,Boyanovsky:2018fxl,Bowen2024} systems where state tomography, even in principle, is inaccessible. 
\par
The frequency-domain master equation developed here can be a powerful analytical and simulation tool in and of itself.  It is known that both the simulation time and size of the propagator grow exponentially with memory length, making it challenging to recover long-time dynamics in the time domain \cite{Vega2017}. Leveraging extensive tools developed for Fourier analysis, simulating the frequency-domain propagator for principal frequency components can be significantly more efficient in terms of computational time and resources. 
\par
Interestingly, our results also provide a means to gauge the effect of erasing state memory as manifested in the positions of the non-Markovian spectral features calculated using frequency-domain and Born-Redfield QMEs. For a system coupled to a bath with squeezed correlations, we find that the frequency-domain master equation preserves positivity of the state, unlike the Born-Redfield description. While our findings are empirical at this stage, both these aspects could be tantalizing indicators of a mechanism for positivity restoration by introducing corrections in the unitary sector of the master equation \cite{Farina2019,Timofeev2022} but now guided by the spectral diagnostic. It will also be interesting to generalize frequency-domain open system descriptions to situations where multi-time and even out-of-time ordered \cite{Syzranov2018}, and not just two-time, correlations are important. {\color{black} Further, the FD-QME may be generalized to higher-order master equations, potentially allowing it to go beyond even the quasi-static bath approximation.}
%
%
\acknowledgments{It is a pleasure to thank Zhihao Xiao and Brenden Bowen for useful discussions. We would like to particularly acknowledge Le Hu and Rahul Trivedi for important pointers regarding positivity breaking in the presence of bath correlations. This work was supported by the National Science Foundation under grant DMR-2047357 and U.S. Department of Energy under grants DE-SC0019515 and DE-SC0020360.}
%
%
%
\appendix
\section{Superoperator notation and algebra}
\label{apdx:Superoperator_notation_and_algebra}
%

The standard evolution of state-vectors, subject to a time-dependent Hamiltonian, is described by the Schr\"{o}dinger equation, 
\begin{IEEEeqnarray}{rCl}
    \partial_t\ket{\psi(t)}
    &=&
    -iH(t)\ket{\psi(t)},
    \startsub
\end{IEEEeqnarray}
resulting in a general solution given by,
\begin{IEEEeqnarray}{rCl}
    \ket{\psi(t)}
    &=&
    T e^{-i\int_0^t\dd{s}H(s)}\ket{\psi(0)}.
    \sub \label{eq:Schrod_Soln}
\end{IEEEeqnarray}
For this same Hamiltonian, the density matrix evolution is described by the Liouville von-Neumann equation, 
\begin{IEEEeqnarray}{rCl}
    \partial_t\rho(t)
    &=&
    -i\qty[H(t),\rho(t)].
    \label{eq:LvN_Gen}
\end{IEEEeqnarray}
In order to recast a solution in a form similar to that in Eq.~(\ref{eq:Schrod_Soln}), we can define a higher rank object which acts on operators: a `superoperator.' Throughout this paper, superoperators are denoted by capital, script characters. How they act on an operator is often indicated by a `dot'. For example, introducing the Liouvillian superoperator
\begin{IEEEeqnarray}{rCl}
    \SO{L}(t)
    &=&
    -i\qty[H(t), \>\vdot],
\end{IEEEeqnarray}
allows us to rewrite Eq.~(\ref{eq:LvN_Gen}) as
\begin{IEEEeqnarray}{rCl}
    \partial_t\rho(t)
    &=&
    \SO{L}(t)\rho(t)
    \startsub \label{eq:DM_Evol_Gen}\\
    \implies \rho(t)
    &=&
    Te^{\int_0^t\dd{s}\SO{L}(s)}\rho(0).
    \sub
\end{IEEEeqnarray}
Superoperators can act on each other, and much like operators, do not generally commute \cite{Carmichael2},
\begin{IEEEeqnarray}{rCl}
    \qty(A \vdot B)\qty(C\vdot C)
    &=&
    AC\vdot CB
    \startsub\\
    \qty(C\vdot C)\qty(A\vdot B)
    &=&
    CA\vdot BC
    \sub
\end{IEEEeqnarray}
Note that, unlike closed systems, superoperators can describe the more general, non-unitary evolution that arises in open quantum systems as well as describe the evolution of general quantum states such as mixed states. For example, the most common superoperator to appear in quantum master equations is the Lindblad-form superoperator, given by
\begin{IEEEeqnarray}{rCl}
    \SO{D}[O]\vdot
    &=&
    2O\vdot O^\dag - O^\dag O\vdot -\vdot O^\dag O.
\end{IEEEeqnarray}
\par
We now generalize the notion of rotating frames in terms of superoperators. Starting with Eq.~(\ref{eq:DM_Evol_Gen}), we then calculate the evolution of $\rho(t)$, in a frame rotating with respect to $-\SO{L}_0$,
\begin{IEEEeqnarray}{rCl}
    \partial_t\qty[e^{-\SO{L}_0t}\rho(t)]
    &=&
    e^{-\SO{L}_0t}\partial_t\rho(t)-\SO{L}_0e^{-\SO{L}_0t}\rho(t)
    \nonumber \\
    &=&
    e^{-\SO{L}_0t}\SO{L}\rho(t)-\SO{L}_0e^{-\SO{L}_0t}\rho(t).
\end{IEEEeqnarray}
Then defining 
\begin{IEEEeqnarray}{rCl}
    \rho_I(t)
    &\equiv&
    e^{-\SO{L}_0t}\rho(t)
    \startsub \\
    \SO{L}_I(t)
    &\equiv&
    e^{-\SO{L}_0t}\SO{L}e^{\SO{L}_0t},
    \sub
\end{IEEEeqnarray}
we find
\begin{IEEEeqnarray}{rCl}
    \partial_t\rho_I(t)
    &=&
    \qty(\SO{L}_I(t)-\SO{L}_0)\rho_I(t).
\end{IEEEeqnarray}
Notice that this is exactly analogous to unitary evolution for a state vector in a rotating frame. If we then break the full Liouvillian into free and interaction parts, $\SO{L}=\SO{L}_0+\SO{V}$, we find the superoperator equivalent of the interaction picture, 
\begin{IEEEeqnarray}{rCl}
    \partial_t\rho_I(t)
    &=&
    \SO{V}_I(t)\rho_I(t).
\end{IEEEeqnarray}

Operators on an $N$-dimensional Hilbert space are represented as $N\times N$ square matrices. Similarly, superoperators can be represented as $N^2\times N^2$ matrices acting on an $N^2$-dimensional Liouvillian space. In this space, operators and density matrices are expressed as $N^2$-vectors in a straightforward way,
\begin{IEEEeqnarray}{rCl}
    \kket{O}
    &=&
    \mqty(
        O_{11} &
        \dots &
        O_{1N} &
        O_{21} &
        \dots &
        O_{NN}
    )^\T. \hspace{10pt}
\end{IEEEeqnarray}
Likewise, we may define the dual vector corresponding to an operator,  $\bbra{O}\equiv\kket{O}^\dag$,
\begin{IEEEeqnarray}{rCl}
    \bbra{O}
    &=&
    \mqty(
        O_{11}^* &
        \ldots &
        O_{1N}^* &
        O_{21}^* &
        \ldots &
        O_{NN}^*
    ),
\end{IEEEeqnarray}
which leads naturally to the Hilbert-Schmidt inner product in Liouvillian space, 
\begin{IEEEeqnarray}{rCl}
    \SOev{A}{B}
    &=&
    \Tr[A^\dag B].
    \label{eq:HS_Inner_Product}
\end{IEEEeqnarray}
%

%
%

\section{System dynamics from Born-Redfield master equation}
\label{apdx:SystemDynamicsFromBornRedfield}


It is instructive to compare the spectrum derived using the frequency-domain master equation to that obtained using the widely used Born-Redfield master equation \cite{BreuerPetruccione}, which is a non-Markovian but time-local map. We derive it by starting from Eq.~(\ref{eq:BornIP}) and making one further approximation: we assume that bath correlation timescales are much shorter than system timescales. Doing so, justifies the replacement $\rho_{S, I}(s)\to\rho_{S, I}(t)$, leading to a time-local equation for $\rho_{I}(t)$, 
\begin{IEEEeqnarray}{rCl}
    \partial_t\rho_{I}(t)
    &=&
    \int_0^tds\Tr_B[\>\SO{V}_I(t)\SO{V}_I(s)\rho_{I}(t)\rho_B^{ss}]. 
    \label{eq:1stMarkIP}
\end{IEEEeqnarray}
This is commonly done in tandem with a substitution of $s\to t-\tau$ and changing the bounds of integration to $[0, \infty)$. This step is equivalent to assuming a delta-correlated bath, as opposed to simply a `fast' one, and yields the well-known Born-Markov master equation. Transforming Eq.~(\ref{eq:1stMarkIP}) back to the lab frame we find
\begin{IEEEeqnarray}{rCl}
    \partial_t\rho(t)
    &=&
    \SO{L}_S \rho(t)+ 
    \int_0^tds\Tr_B[\>\SO{V}\SO{V}_I(s-t)\rho(t)\rho_B^{ss}]. 
    \label{eq:BRME_DM} 
    \nonumber \\
    &\equiv&
    \qty(\SO{L}_S+\SO{L}^{(2)}_{\text{BR}}(t))\rho(t) 
    \label{eq:BR_ME_Thermal}
\end{IEEEeqnarray}
Notice the relationship between this second-order Liouvillian and the memory kernel from Eq.~(\ref{eq:NLME_Vec}). It may be expressed concisely as
\begin{IEEEeqnarray}{rCl}
    \SO{L}^{(2)}_{\text{BR}}(t)
    &=&
    \int_0^t \dd{s}\SO{K}^{(2)}(t-s)e^{-\SO{L}_S(t-s)}.
    \label{eq:BR_from_Kernel}
\end{IEEEeqnarray}
Having already calculated the kernel in Eq.~(\ref{eq:Kernel_Therm}), we simply evaluate the above integral and obtain, 
\begin{IEEEeqnarray}{rCl}
    \SO{L}^{(2)}_{\text{BR}}(t)
    &=&
    -i\delta_{\text{eff}}(t)\qty[\qty(\bar{N}+1)\sigma_+\sigma_--\bar{N}\sigma_-\sigma_+, \>\vdot] 
    \nonumber \\
    &&
    +\gamma_{\text{eff}}(t)\qty(\qty(\bar{N}+1)\SO{D}[\sigma_-]\vdot+\bar{N}\SO{D}[\sigma_+]\vdot), \hspace{5pt}
    \label{eq:BR_Therm_2nd_Order_Dissipators}
\end{IEEEeqnarray}
where we have defined the effective frequency shift and decay rate of the qubit,
\begin{IEEEeqnarray}{rCl}
    \delta_{\text{eff}}(t)
    &=&
    \frac{ g^2}{\Delta^2+\kappa^2}
    \qty(\Delta\qty(1\hspace{-2pt}-\hspace{-2pt}e^{-\kappa t}\cos \Delta t) -\kappa e^{-\kappa t}\sin \Delta t),
    \nonumber \\ 
    \label{eq:BR_Therm_Rates} \startsub \\
    \gamma_{\text{eff}}(t)
    &=&
    \frac{ g^2}{\Delta^2+\kappa^2}
    \qty(\kappa\qty(1\hspace{-2pt}-\hspace{-2pt}e^{-\kappa t}\cos \Delta t) +\Delta e^{-\kappa t}\sin \Delta t). 
    \nonumber \\
    \sub 
\end{IEEEeqnarray}
\begin{figure}[t!]
    \centering
    \includegraphics[width=\columnwidth]{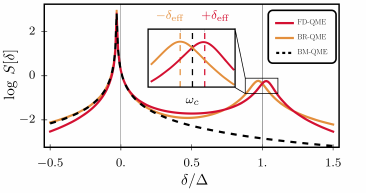}

    \caption{
        Steady state emission spectrum of a qubit interacting with a thermal bath, calculated using time-non-local solution in Eq.~(\ref{eq:FD_Therm_Spec_Kernel}) (red) and the time-local solution in Eq.~(\ref{eq:BR_Spec}) (orange) respectively. The inset shows that the location of the side peak differs between the two solutions, appearing on opposite sides of the cavity frequency. Parameters: $\omega_q/g=2\times 10^5$, $\Delta/g=10$, $\kappa/g=0.5$, $\bar{N}=1.0$.
    }
    \label{fig:Compare_Spectra}
\end{figure}

We now use the Born-Redfield master equation to calculate the two-time correlator, $\ev{\sigma_+(\tau)\sigma_-}_{ss}$. Normalized by the excited state population, we find
\begin{IEEEeqnarray}{rCl}
    G(\tau)
    &=&
    \lim_{t\to\infty}\frac{\ev{\sigma_+(t+\tau)\sigma_-(t)}}{\ev{\sigma_+(t)\sigma_-(t)}} 
    \nonumber \\
    &=&
    \exp\Bigg(i\omega_q\tau-\frac{\qty(2\bar{N}+1) g^2\tau}{\qty(\kappa+i\Delta)} 
    \nonumber \\
    && \hspace{24pt}
    +\frac{\qty(2\bar{N}+1) g^2\qty(1-e^{-\qty(\kappa+i\Delta)\tau})}{\qty(\kappa+i\Delta)^2}\Bigg).
\end{IEEEeqnarray}
The spectrum is obtained by Fourier transforming $G(\tau)$; however, the nested exponential makes this difficult. Instead, we obtain an approximate solution by expanding the outer exponent to first order in $g^2$. Transforming this to the frequency domain, we find
\begin{IEEEeqnarray}{rCl}
    S_{\text{BR}}[\delta]
    &\approx&
    \frac{1}{\pi}\frac{\gamma_{\text{eff}}}{\qty(\delta-\delta_{\text{eff}})^2+\gamma_{\text{eff}}^2} 
    \nonumber \\
    &&
    +\frac{1}{\pi}
    \frac{\delta_{\text{eff}}\Delta-\gamma_{\text{eff}}\kappa}{\Delta^2+\kappa^2}
    \frac{
        \gamma_{\text{eff}}+\kappa
    }{
        \qty(\delta-\delta_{\text{eff}}+\Delta)^2+(\gamma_{\text{eff}}+\kappa)^2
    }
    \nonumber \\
    &&
    +\frac{1}{\pi}
    \frac{2\sqrt{\delta_{\text{eff}}\Delta\gamma_{\text{eff}}\kappa}}{\Delta^2+\kappa^2}
    \frac{
        \delta-\delta_{\text{eff}}+\Delta
    }{
        \qty(\delta-\delta_{\text{eff}}+\Delta)^2+(\gamma_{\text{eff}}+\kappa)^2
    }.
    \nonumber \\
    \label{eq:BR_Spec}
\end{IEEEeqnarray}

As in Sec. \ref{sec:ExampleJCI}, we see a central Lorentzian, given by the first-order term, and a side peak near the cavity frequency, given by the second-order term. Notice in Fig. \ref{fig:Compare_Spectra}, the position of the side peak differs between the two curves. From Eq.~(\ref{eq:BR_Spec}) it is clear that the central and side peak are separated by $\Delta$. If the spectrum is calculated using the method described in Sec. \ref{subsec:SpetraFromCorrelatorsInFrequencyDomain}, we instead see a separation of roughly $\Delta+2\delta_{\text{eff}}$, which is consistent with simulations of the full system.
%
%

\section{System spectrum with squeezed bath}
\label{apdx:SqueezedBathFullDerivation}


Consider the full Hamiltonian given in Eq.~(\ref{eq:Squeeze_Hamiltonian}),
\begin{IEEEeqnarray}{rCl}
    H
    &=&
    \underbrace{\frac{\Delta_q}{2}\sigma_z}_{H_S}+
    \underbrace{\Delta_c a^\dag a+\frac{r}{2}\qty(a^2+a^{\dag 2})}_{H_B}+
    \underbrace{ g\qty(\sigma_-a^\dag + \sigma_+ a)}_V 
    \nonumber \\
\end{IEEEeqnarray}
It is shown in \cite{Stern22} how a circuit QED system may be engineered to have a Hamiltonian of this form. The cavity is also subject to Markovian loss modeled by a Lindblad dissipator $\SO{D}_B=\kappa\SO{D}[a]\vdot$. In order to express the interaction Liouvillian, $\SO{V}=-i\qty[V,\> \vdot]$, in the interaction picture defined with respect to $\SO{L}_0=-i\qty[H_S+H_B, \>\vdot]+\SO{D}_B$, it is useful to first move to the eigenbasis of $H_B$. To this end, we apply a Bogoliubov transformation to the cavity operators,
\begin{IEEEeqnarray}{rCl}
    \~a
    &=&
    a\cosh{\zeta}+a^\dag\sinh{\zeta},
    \label{eq:Bogo}
\end{IEEEeqnarray}
leading to
\begin{IEEEeqnarray}{rCl}
    H_B
    &=&
    \widetilde{\Delta}_c\~a^\dag\~a,
\end{IEEEeqnarray}
where
\begin{IEEEeqnarray}{rCl}
    \zeta
    &=&
    \frac12\tanh^{-1}\qty(\frac{r}{\Delta_c}), 
    \startsub \\
    \widetilde{\Delta}_c
    &=&
    \sqrt{\Delta_c^2-r^2}. 
    \sub \label{eq:BathEffectiveFreq}
\end{IEEEeqnarray}
In the Bogoliubov basis, the interaction Hamiltonian reads
\begin{IEEEeqnarray}{rCl}
    V
    &=&
     g_1\qty(\sigma_-\~a^\dag + \sigma_+ \~a)+
     g_2\qty(\sigma_-\~a + \sigma_+ \~a^\dag),
\end{IEEEeqnarray}
where, for simplicity and generality, we define two independent effective couplings $ g_1\equiv g\cosh(\zeta)$ and ${g_2\equiv- g\sinh(\zeta)}$. 
\par
Finally, we express the dissipator, $\SO{D}_B$, in the Bogoliubov basis using the definition of $\SO{D}[O]\vdot$ and Eq.~(\ref{eq:Bogo}),
\begin{IEEEeqnarray}{rCl}
    \SO{D}_B
    &=&
    \kappa\Big(\cosh^2(\zeta)\SO{D}[\~a]\vdot+\sinh^2(\zeta)\SO{D}[\~a^\dag]\vdot 
    \nonumber \\
    &&\hspace{20pt}
    -\frac12\sinh(2\zeta)\qty(\SO{S}[\~a]\vdot+\SO{S}[\~a^\dag]\vdot)\Big),
    \label{eq:DissBogo}
\end{IEEEeqnarray}
where we define $\SO{S}[O]\vdot=2O\vdot O-O^2\vdot-\vdot O^2$. The form of dissipator in Eq.~(\ref{eq:DissBogo}) shows that a squeezed cavity subject to loss due to vacuum fluctuations can be equivalently represented as an unsqueezed cavity subject to dissipation due to squeezed thermal fluctuations, with the effective temperature of the fluctuations determined by the squeezing strength. We now express the cavity dissipator entirely in terms of steady-state correlators as
\begin{IEEEeqnarray}{rCl}
    \SO{D}_B
    &=&
    \kappa\qty((\bar{N}+1)\SO{D}[\~a]\vdot+\bar{N}\SO{D}[\~a^\dag]\vdot) 
    \nonumber \\
    &&
    -(\kappa+i\widetilde{\Delta}_c)\bar{M}\qty(\SO{S}[\~a]\vdot+\SO{S}[\~a^\dag]\vdot).
\end{IEEEeqnarray}
where
\begin{IEEEeqnarray}{rCl}
     & & \bar{N}\equiv\ev{\~a^\dag\~a}_{ss} = \sinh^2(\zeta),
     \startsub \\
     & & \bar{M}\equiv\ev{\~a^2}_{ss} = \frac{\kappa\sinh(2\zeta)}{2\qty(\kappa+i\widetilde{\Delta}_c)}.
     \sub
\end{IEEEeqnarray}
\par
For ease of calculation, next we express the interaction Liouvillian in terms of a matrix product,
\begin{IEEEeqnarray}{rCl}
    \SO{V}
    &=&
    -i\vb{\Sigma}^\T\mat{G}\vb{A}=-i\vb{A}^\T\mat{G}\vb{\Sigma}, 
    \label{eq:InteractionMatrixForm}
\end{IEEEeqnarray}
where we have defined the following vectors of superoperators,
\begin{IEEEeqnarray}{rCl}
    \vb{\Sigma}
    &\equiv&
    \mqty((\sigma_+\vdot) & (\sigma_-\vdot) & (\vdot\sigma_-) & (\vdot\sigma_+))^\T 
    \nonumber \\
    \vb{A}
    &\equiv&
    \mqty((\~a\vdot) & (\~a^\dag\vdot) & (\vdot \~a^\dag) & (\vdot \~a))^\T, 
    \nonumber
\end{IEEEeqnarray}
and the coupling matrix,
\begin{IEEEeqnarray}{rCl}
    \mat{G}
    &\equiv&
    \mqty(
         g_1   &  g_2 &           &           \\
         g_2   &  g_1 &           &           \\
                    &           &- g_1 &- g_2 \\
                    &           &- g_2 &- g_1
        ). 
    \nonumber
\end{IEEEeqnarray}
It is important to note that these vectors are not actual quantum objects; they are simply matrices with superoperator elements and follow standard matrix operations. Whether the elements act through scalar multiplication or superoperator action should be clear by the context. We may write a similar expression in the interaction picture,
\begin{IEEEeqnarray}{rCl}
    \SO{V}_I(t)
    &=&
    -i\vb{\Sigma}_I^\T(t)\mat{G}\vb{A}_I(t)=-i\vb{A}_I^\T(t)\mat{G}\vb{\Sigma}_I(t), \label{eq:InteractionIPMatrixForm}
\end{IEEEeqnarray}
where 
\begin{IEEEeqnarray}{rCl}
    \partial_t\vb{A}_I(t)
    &=&
    -\qty[\SO{L}_B, \vb{A}_I(t)], 
    \startsub \\
    \partial_t\vb{\Sigma}_I(t)
    &=&
    -\qty[\SO{L}_S, \vb{\Sigma}_I(t)]. 
    \sub
\end{IEEEeqnarray}
Here the commutators are evaluated element-wise on the vectors. Ignoring the system part for the moment, we rewrite this as a matrix equation,
\begin{IEEEeqnarray}{rCl}
    \partial_t\vb{A}_I(t)
    &=&
    -\mat{M}\vb{A}_I(t),
\end{IEEEeqnarray}
where, after some tedious algebra, we find
\begin{widetext}
\begin{IEEEeqnarray}{rCl}
    \mat{M}
    &=&
    \mqty(
        i\widetilde{\Delta}_c+(2\bar{N}+1)\kappa 
        & 
        -2(\kappa+i\widetilde{\Delta}_c)\bar{M} 
        & 
        2(\kappa+i\widetilde{\Delta}_c)\bar{M} 
        & 
        -2\bar{N}\kappa 
        \\
        2(\kappa+i\widetilde{\Delta}_c)\bar{M} 
        & 
        -i\widetilde{\Delta}_c-(2\bar{N}+1)\kappa 
        & 
        2(\bar{N}+1)\kappa 
        & 
        -2(\kappa+i\widetilde{\Delta}_c)\bar{M} 
        \\
        2(\kappa+i\widetilde{\Delta}_c)\bar{M} 
        & 
        -2\bar{N}\kappa 
        &
        -i\widetilde{\Delta}_c+(2\bar{N}+1)\kappa 
        & 
        -2(\kappa+i\widetilde{\Delta}_c)\bar{M}
        \\
        2(\bar{N}+1)\kappa 
        & 
        -2(\kappa+i\widetilde{\Delta}_c)\bar{M} 
        & 
        2(\kappa+i\widetilde{\Delta}_c)\bar{M} 
        & 
        i\widetilde{\Delta}_c-(2\bar{N}+1)\kappa
),
\end{IEEEeqnarray}
with the solution for cavity superoperators simply given by,
\begin{IEEEeqnarray}{rCl}
    \vb{A}_I(t)
    &=&
    e^{-\mat{M}t}\vb{A}.
\end{IEEEeqnarray}
We can use the same procedure to find the evolution of qubit superoperators, $\vb{\Sigma}_I(t)$; however, the process is trivial since the system has only a Hamiltonian evolution. Expressed in terms of these matrices, the memory kernel, $\SO{K}^{(2)}(t)$, can then be written as
\begin{IEEEeqnarray}{rCl}
    \SO{K}^{(2)}(t)
    &=&
    \Tr_B[\SO{V}\SO{V}_I(-t)\rho_B^{ss}]e^{\SO{L}_S t} 
    \nonumber \\
    &=&
    -\Tr_B[\qty(\vb{\Sigma}^\T\mat{G}\vb{A})\qty(\vb{A}^\T_I(-t)\mat{G}\vb{\Sigma}_I(-t))\rho_B^{ss}]e^{\SO{L}_S t} 
    \nonumber \\
    &=&
    -\Tr_B[\vb{\Sigma}^\T\mat{G}
    \qty(\vb{A}\vb{A}^\T\rho_B^{ss})
    (e^{\mat{M}\>t})^\T\mat{G}\vb{\Sigma}_I(-t)]e^{\SO{L}_S t}. 
    \label{eq:BR2ndOrderLiouvillian1}
\end{IEEEeqnarray}
In the last line, we have exploited the fact that $\rho_B^{ss}$ commutes with everything except the $\vb{A}$ vectors. The partial trace may now be taken (element-wise), resulting in a correlator matrix,
\begin{IEEEeqnarray}{rcl}
    \mat{T}\equiv\Tr_B\qty[\vb{A}\vb{A}^\T\rho^{ss}_B] 
    &=&
    \mqty(
        \bar{M} & \bar{N}+1 & \bar{N}   & \bar{M} \\
        \bar{N} & \bar{M}^* & \bar{M}^* & \bar{N}+1 \\
        \bar{N} & \bar{M}^* & \bar{M}^* & \bar{N}+1 \\
        \bar{M} & \bar{N}+1 & \bar{N}   & \bar{M}
    ).
\end{IEEEeqnarray}
Now, we can express the memory kernel entirely in terms of system superoperators, 
\begin{IEEEeqnarray}{rCl}
    \SO{K}^{(2)}(t)
    &=&
    -\vb{\Sigma}^\T\mat{G}\mat{T}(e^{\mat{M}\>t})^\T\mat{G}\vb{\Sigma}_I(-t)e^{\SO{L}_S t}, \label{eq:BR2ndOrderLiouvillian2}
\end{IEEEeqnarray}
which, though cumbersome, is amenable to analytical evaluation. Writing the memory kernel as a matrix, as in Eq.~(\ref{eq:Kernel_Therm}), 
\begin{IEEEeqnarray}{rCl}
    \SO{K}^{(2)}(t)
    &=&
    \smqty(
        \SO{K}_{11}(t)  &                   &                   &-\SO{K}_{44}(t)    \\
                        &\SO{K}_{22}(t)     &\SO{K}_{32}^*(t)     &                   \\
                        &\SO{K}_{32}(t)   &\SO{K}_{22}^*(t)   &                   \\          
        -\SO{K}_{11}(t) &                   &                   &\SO{K}_{44}(t)     \\
    ),
\end{IEEEeqnarray}
we find,
\begin{IEEEeqnarray}{rCl}
    \vspace{6pt}\SO{K}_{11}(t)
    &=&
    -e^{-\kappa t}\qty[
         g_1 g_2\bar{M}\qty(e^{-i\widetilde{\Delta} t}+e^{i\widetilde\Sigma t})+
         g_1^2\bar{N}\cos\widetilde{\Delta} t+
         g_2^2(\bar{N}+1)\cos\widetilde\Sigma t+
        \text{c.c.}
    ],
    \label{eq:Sq_Kernel_Time_Domain}\startsub \\
    \vspace{6pt}\SO{K}_{22}(t)
    &=&
    -e^{-\kappa t}\qty[
        \qty(2 g_1 g_2\bar{M}+ g_1^2(2\bar{N}+1))e^{i\~\omega_c t}+
        \qty(2 g_1 g_2\bar{M}^*+ g_2^2(2\bar{N}+1))e^{-i\~\omega_c t}
    ],
    \sub \\
    \vspace{6pt}\SO{K}_{32}(t)
    &=&
    e^{-\kappa t}\qty[
        \qty(2 g_1^2\bar{M}^*+ g_1 g_2(2\bar{N}+1))e^{-i\~\omega_c t}+
        \qty(2 g_2^2\bar{M}+ g_1 g_2(2\bar{N}+1))e^{i\~\omega_c t}
    ],\hspace{20pt}
    \sub \\
    \SO{K}_{44}(t)
    &=&
    -e^{-\kappa t}\qty[
         g_1 g_2\bar{M}\qty(e^{-i\widetilde{\Delta} t}+e^{i\widetilde\Sigma t})+
         g_1^2(\bar{N}+1)\cos\widetilde{\Delta} t+ g_2^2\bar{N}\cos\widetilde\Sigma t+
        \text{c.c.}
    ]. 
    \sub
\end{IEEEeqnarray}
Here we have defined the effective qubit-cavity difference frequency, $\widetilde{\Delta}\equiv\Delta_q-\widetilde{\Delta}_c$, and sum frequency, $\widetilde\Sigma\equiv\Delta_q+\widetilde{\Delta}_c$. Note that as compared to the case of thermal cavity, Eq.~(\ref{eq:Therm_Kernel_Time_Domain}), the additional term due to squeezed bath leads to a $\SO{K}_{32}(t)$ term signifying the squeezing-induced modification of qubit coherences. 
\par
The time-dependence of functions in Eq.~(\ref{eq:Sq_Kernel_Time_Domain}) is comprised of simple exponentials, making it easy to transform them into the frequency domain. Following the procedure given in Sec.~\ref{subsec:SpetraFromCorrelatorsInFrequencyDomain}, we calculate the steady-state of the system. As in the thermal case, we find it to be diagonal, but now with
\begin{IEEEeqnarray}{rCl}
    &&
    \bra{0}\rho_{ss}\ket{0}=
    \nonumber \\
    &&
    \frac{
        \kappa\qty[
            (\bar{N}+1)g_1^2\qty(\kappa^2+\widetilde\Sigma^2)
            +
            \bar{N}g_2^2\qty(\kappa^2+\widetilde\Delta^2)
        ]
        +
        g_1g_2\qty[
            \Re[\bar{M}]\kappa\qty(2\kappa^2+\widetilde\Delta^2+\widetilde\Sigma^2)
            +
            \qty(\widetilde\Sigma-\widetilde\Delta) \Im[\bar{M}]\qty(\kappa^2+\widetilde\Sigma\widetilde\Delta)
        ]
    }{
        (2\bar{N}+1)\kappa\qty[
            g_1^2\qty(\kappa^2+\widetilde\Sigma^2)
            +
            g_2^2\qty(\kappa^2+\widetilde\Delta^2)
        ]
        +
        2g_1g_2\qty[
            \Re[\bar{M}]\kappa\qty(2\kappa^2+\widetilde\Delta^2+\widetilde\Sigma^2)
            +
            \qty(\widetilde\Sigma-\widetilde\Delta) \Im[\bar{M}]\qty(\kappa^2+\widetilde\Sigma\widetilde\Delta)
        ]
    }.
    \nonumber \\
\end{IEEEeqnarray}
Notice that in the absence of squeezing, $\bar{M}=\bar{N}=0$ and $g_2=0$, the vacuum limit of Eq.~(\ref{eq:Thermal_SS_DM}) is recovered. Now using this to calculate the steady-state emission spectrum given in Eq.~(\ref{eq:Spec_Fin}), we find
\begin{IEEEeqnarray}{rCl}
    S[\delta]
    &=& 
    -\frac{1}{\pi}\qty(
        \Re\SO{K}_{22}[\delta]-
        \frac{\Re\bar{\SO{K}}_{32}[\delta]\Re\SO{K}_{22}[-\delta]-\Im\bar{\SO{K}}_{32}[\delta]\qty(\delta+2\Delta_q+\Im\SO{K}_{22}[-\delta])}{\qty(\Re\SO{K}_{22}[-\delta])^2+\qty(\delta+2\Delta_q+\Im\SO{K}_{22}[-\delta])^2}
    ) 
    \nonumber \\
    &&\hspace{10pt}
    \times\left[
        \qty(\Re\SO{K}_{22}[\delta]-
        \frac{\Re\bar{\SO{K}}_{32}[\delta]\Re\SO{K}_{22}[-\delta]-\Im\bar{\SO{K}}_{32}[\delta]\qty(\delta+2\Delta_q+\Im\SO{K}_{22}[-\delta])}{\qty(\Re\SO{K}_{22}[-\delta])^2+\qty(\delta+2\Delta_q+\Im\SO{K}_{22}[-\delta])^2}
        )^2\right.
    \nonumber \\
    &&
        \hspace{20pt}\left.+
        \qty(
            \delta -\Im\SO{K}_{22}[\delta]+
            \frac{
                \Im\bar{\SO{K}}_{32}[\delta]\Re\SO{K}_{22}[-\delta]+\Re\bar{\SO{K}}_{32}[\delta]\qty(\delta+2\Delta_q+\Im\SO{K}_{22}[-\delta])
            }{
                \qty(\Re\SO{K}_{22}[-\delta])^2+\qty(\delta+2\Delta_q+\Im\SO{K}_{22}[-\delta])^2
            }
        )^2
    \right]^{-1},
    \label{eq:FD_Squeeze_Spec}
\end{IEEEeqnarray}
where we have defined
\begin{IEEEeqnarray}{rCl}
    \bar{\SO{K}}_{32}[\delta]
    &=&
    \SO{K}_{32}[\delta]\SO{K}^*_{32}[\delta]
    \neq
    \qty|\SO{K}_{32}[\delta]|^2.
    \label{eq:K32bar}
\end{IEEEeqnarray}
Note the similarity to Eq.~(\ref{eq:FD_Therm_Spec_Kernel}) but with additional terms proportional to $\bar{\SO{K}}_{32}[\delta]$. One can verify that when $r\to 0$, the spectrum for the thermal case in the zero temperature limit is recovered.

As done in the case of a thermal state in the cavity, we can compare the frequency-domain master equation results for a squeezed cavity with those obtained using the Born-Redfield master equation. Using Eq.~(\ref{eq:BR_from_Kernel}), we find the corresponding second-order Liouvillian,
\begin{IEEEeqnarray}{rCl}
    \SO{L}_{\text{BR}}^{(2)}(t)
    &=&
    \int_0^t\dd{s}\mqty(
        \SO{K}_{11}(s)  &                                   &                               &-\SO{K}_{44}(s)\\
                        &e^{-i\Delta_q s}\SO{K}_{22}(s)     &e^{i\Delta_q s}\SO{K}_{32}(s)  &               \\
                        &e^{-i\Delta_q s}\SO{K}_{32}^*(s)   &e^{i\Delta_q s}\SO{K}_{22}^*(s)&               \\
        -\SO{K}_{11}(s) &                                   &                               &\SO{K}_{44}(s) \\
    ).
\end{IEEEeqnarray}
\end{widetext}
\mbox{~}
After evaluating the integral, we may express this in terms of commutators and dissipators,
\begin{IEEEeqnarray}{rCl}
    \SO{L}^{(2)}_{\text{BR}}
    &=&
    -i\qty[\delta_{+-}(t)\sigma_+\sigma_--\delta_{-+}(t)\sigma_-\sigma_+, \>\vdot] 
    \nonumber \\
    &&
    +\qty(\gamma_{-+}(t)\SO{D}[\sigma_-]\vdot+\gamma_{+-}(t)\SO{D}[\sigma_+]\vdot) 
    \nonumber \\
    &&
    +\qty(\gamma_{--}(t)\SO{S}[\sigma_-]\vdot+\gamma_{++}(t)\SO{S}[\sigma_+]\vdot).
    \label{eq:BR_Sq_2nd_Order_Dissipators}
\end{IEEEeqnarray}
As evident, the form is similar to that in Eq.~(\ref{eq:BR_Therm_2nd_Order_Dissipators}), but with the addition of two squeezing dissipators, $\SO{S}[\sigma_\pm]\vdot$, with the time-dependent rates given by
\begin{IEEEeqnarray}{rCl}
    \gamma_{-+}(t)
    &=&
    \Re\qty[
        \qty(\qty(\bar{N}+1) g_1^2+\bar{M} g_1 g_2)
        \frac{1-e^{-(\kappa+i\widetilde{\Delta})t}}{\kappa+i\widetilde{\Delta}}
    ],
    \nonumber \\
    \startsub \\
    \gamma_{+-}(t)
    &=&
    \Re\qty[
        \qty(\bar{N} g_1^2+\bar{M} g_1 g_2)
        \frac{1-e^{-(\kappa+i\widetilde{\Delta})t}}{\kappa+i\widetilde{\Delta}}
    ],
    \nonumber \\
    \sub \\
    \gamma_{--}(t) 
    &=& 
    \gamma_{++}^{*}(t) \nonumber\\
    &=&
    \qty(\bar{M} g_2^2+\frac{2\bar{N}+1}{2} g_1 g_2)
    \qty(\frac{1-e^{-(\kappa+i\widetilde{\Delta})t}}{\kappa+i\widetilde{\Delta}}).
    \nonumber \\
    \sub
\end{IEEEeqnarray}
Here we have ignored all sum-frequency terms for brevity; they may still be included in further calculations but have a muted effect on the full dynamics. The commutator terms will result in time-dependent frequency shifts, $\delta_{+-}(t)$ and $\delta_{-+}(t)$, given by imaginary parts of $\gamma_{-+}$ and $\gamma_{+-}$, respectively. As before, if the squeezing strength is set to zero, $r\to 0$, the results in Eq.~(\ref{eq:BR_Therm_Rates}) are recovered. 
\par
\begin{figure}[t!]
    \centering
    \includegraphics[width=\columnwidth]{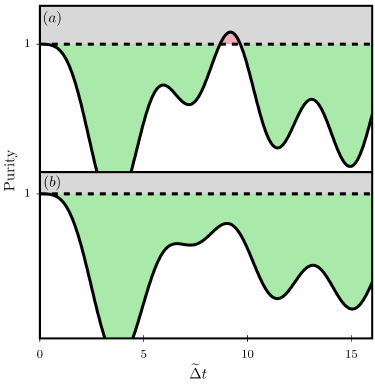}

    \caption{
        Qubit purity, $\Tr[\rho^2]$, as a function of time calculated using (a) the Born-Redfield master equation and (b) the frequency-domain master equation. The gray region corresponds to purity a greater than $1$ and is equivalent to a negative eigenvalue. As shown in red, the BR-QME does show a transient negative eigenvalue, indicating that it fails to preserve positivity in this regime. The FD-QME, however, remains below unit-purity at all times and preserving positivity at all times. Parameters: $\Delta_q/g=200$, $\Delta_c/g=120$, $\~\Delta_c/g\approx 34$, $\kappa/g=10$.
        }
    \label{fig:SqueezedPositivity}
\end{figure}

It is also important to note that the Born-Redfield master equation fails to preserve the positivity of $\rho(t)$ for large values of $r$. As shown in Fig.~\ref{fig:SqueezedPositivity}, the system initialized in $\ket{\sigma_y=-1}$, has a purity greater than one for a short time leading to a negative eigenvalue in the system density matrix. Upon numerically transforming the density matrix obtained from the frequency-domain master equation back to the time domain, however, we see that $\Tr[\rho^2]\le1$ for all times and therefore it preserves positivity.
%
%
%
%
%
\clearpage

\end{document}